\documentclass{imamat}

\usepackage{amsmath} 
\usepackage{amsthm} 
\usepackage{graphicx} 

\begin{document}

\title{The inverse problem for collisionless plasma equilibria}

\shorttitle{The inverse problem for collisionless plasma equilibria} 
\shortauthorlist{O.Allanson, S. Troscheit \& T. Neukirch} 

\author{
\name{Oliver Allanson$^*$}
\address{Solar \& Magnetospheric Theory Group, School of Mathematics \& Statistics, \\University of St Andrews, St Andrews, KY16 9SS, UK.}
\address{Space and Atmospheric Electricity Group, Department of Meteorology, \\University of Reading, Reading, RG6 6BB, UK.}\email{$^*$Corresponding author: o.allanson@reading.ac.uk}
\name{Sascha Troscheit}
\address{Analysis Group, School of Mathematics \& Statistics, University of St Andrews, \\St Andrews, KY16 9SS, UK.}
\address{Department of Pure Mathematics, Faculty of Mathematics, University of Waterloo,\\ Waterloo, N2L 3G1, Canada.}\email{stroscheit@uwaterloo.ca}
\and
\name{Thomas Neukirch}
\address{Solar \& Magnetospheric Theory Group, School of Mathematics \& Statistics, University of St Andrews, \\St Andrews, KY16 9SS, UK.}
\email{tn3@st-andrews.ac.uk}
}

\maketitle

\begin{abstract}
  {Vlasov-Maxwell equilibria are described by the self-consistent solutions of the time-independent Maxwell equations for the real-space dynamics of electromagnetic fields, and the Vlasov equation for the phase-space dynamics of particle distributions in a collisionless plasma. These two systems (macroscopic and microscopic) are coupled via the source terms in Maxwell's equations, which are sums of velocity-space `moment' integrals of the particle distribution function. This paper considers `the inverse problem for collisionless equilibria' (IPCE), viz. \emph{``given information regarding the real-space/macroscopic configuration of a specific collisionless plasma equilibrium, what self-consistent equilibrium distributions exist?''} We develop the constants of motion approach to IPCE using the assumptions of a `modified Maxwellian' distribution function, and a strictly neutral and spatially one-dimensional plasma. In such circumstances, IPCE formally reduces to the inversion of Weierstrass transformations \citep{Bilodeau-1962}, such as those transformations that feature in the initial value problem for the heat/diffusion equation. We discuss the various mathematical conditions that a candidate solution of IPCE must satisfy. One method that can be used to invert the Weierstrass transform is expansions in Hermite polynomials. Building on the results of \citet{Allanson-2016JPP}, we establish under what circumstances a solution obtained by these means converges, and allows velocity moments of all orders. Ever since the seminal work by \citet{Bernstein-1957}, on `stationary' electrostatic plasma waves, the necessary quality of non-negativity has been noted as a feature that any candidate solution of IPCE will not \emph{a priori} satisfy. We also discuss this problem in the context of for our formalism, for magnetised plasmas.}
{Insert keyword text here.}
  \\
\end{abstract}

\section{Introduction}
It is estimated that more than $99\%$ of the observable matter in the universe is in the plasma state \citep{Baumjohannbook}. These plasmas are frequently sufficiently hot and/or diffuse that the particles rarely collide, for example the collisional \emph{mean free path} in the solar wind is approximately equal to the distance from the Sun to the Earth \citep{Marsh-2006}. In such circumstances, the plasma dynamics can be accurately modelled without including collisions (e.g. see \citet{Schindlerbook, belmontbook}). Collisionless plasmas behave quite differently to collisional plasmas, and are best modelled using the Vlasov kinetic theory of particle distributions in position-velocity phase space (e.g. see \citet{chenbook, krallbook-1973}), rather than collisional fluid models that operate in position space only (e.g. see \citet{KulsrudMHD, Freidbergbook} for discussions of the Magnetohydrodynamic theory). 

In nature, this difference is made manifest by the critical dependence of the macroscopic dynamics of collisionless plasmas on the velocity space structure of particle distributions \citep{Gary-2005}, and on the short length and time scale physics of collisionless processes (e.g. see \citet{lifshitzkinetic, kulsrudbook, Schindlerbook}). Modern instrumentation is now able to make observations of particle distributions with spatio-temporal resolution on kinetic scales, for example the NASA Multiscale Magnetospheric (MMS) mission \citep{Hesse-2016}, and the ESA candidate mission: Turbulent Heating ObserveR (THOR) \citep{Vaivads-2016}. As such, it is of interest and importance to plasma observers, modellers and theorists to better understand the micro-scale kinetic physics of collisionless plasmas, and in particular how this relates to the macroscopic dynamics, which are typically better understood. 

Collisionless plasma dynamics are frequently modelled using self-consistent particle-in-cell (PIC) codes (e.g. see \citet{birdsallbook}). In both PIC simulations and \emph{ab initio} kinetic theory analyses, a very common approach that is used to model the dynamics is to first set up a Vlasov equilibrium state, then make a perturbation, and next track the subsequent kinetic evolution and the effect on the macroscopic physics (e.g. see \citet{Drake-1977, Daughton-1999}). Due to the technical difficulty in calculating exact Vlasov equilibria, approximate solutions are often used as initial conditions instead of exact solutions (e.g. see \citet{Swisdak-2003, Hesse-2005, Pritchett-2008, Malakit-2010, Aunai-2013, Hesse-2013, Guo-2014, Hesse-2014, Liu-2016}), and it is not always known how far the approxmate solution is from a true equilibrium.  As such, it is of interest to modellers and theorists to better understand the equilibrium states of Vlasov plasmas. However, there is not a well-established `user-friendly' theory that relates the equilibrium and stability properties of the macroscopic real-space description of the plasma to those of the microscopic phase-space description, in the sense of a `one-to-one' correspondence, i.e. `micro$\leftrightarrow$macro'. 

Calculating a kinetic equilibrium given some macroscopic conditions is an example of a non-unique inverse problem: `the inverse problem in collisionless equilibria' (IPCE) (e.g. see \citet{Channell-1976, Mynick-1979a, Harrison-2009POP} for examples of IPCE solutions, and \citet{Allanson-2016JPP} for examples and discussion). This paper considers IPCE applied to spatially one-dimensional (1D) plasmas, and focusses in particular on the mathematical validity of 1D IPCE solutions. 

This paper is structured as follows. In Section 2. we introduce the basic theory of the Vlasov equation and the equations of Vlasov-Maxwell equilibria. We discuss the two distinct approaches to calculating self-consistent collisionless plasma equilibria (`forward' and `inverse'), and previous works on the non-negativity of DFs obtained in IPCE. Section 3. focusses on IPCE in spatially 1D plasmas, outlines parallels between the mathematical formalism of IPCE as presented, and the initial value problem for the heat equation, and introduces the Fourier transform method of formally inverting the equations in this formalism to obtain IPCE solutions. An alternative method to solve IPCE is to use expansions in Hermite polynomials, and this method is outlined in Section 4. One downfall of this method is that the non-negativity of expansions in Hermite polynomials is not \emph{a priori} known. We present general results on the convergence of the Hermite polynomial expansions, and the existence of velocity moments of all orders, and we discuss the problem of non-negativity. In Section 5. we summarise the theory and results presented in this paper.

\section{Basic Theory}
\subsection{The Vlasov Equation}
The Vlasov equation \citep{Vlasov-1968} - also known as the collisionless Boltzmann equation - governs the evolution of the DF for `particle' species $s$, $f_s$, in six-dimensional phase-space, $(\mathbf{x}(t),\mathbf{v}(t))$. Here, $\mathbf{x}$ and $\mathbf{v}=d_t\mathbf{x}$ denote the particle position and velocity, respectively. The Vlasov equation can be used to model the statistical behaviour of collisionless particle distributions for rarefied gases and plasmas, or even to describe the distribution of stars (e.g. see \citet{Henon-1982, Pegoraro-2015} for short introductions to some relevant literature and applications). The Vlasov equation is given in Cartesian geometry by
\begin{equation}
\frac{d}{dt}f_{s}\left(\mathbf{x}(t),\mathbf{v}(t);t\right)=\frac{\partial f_s}{\partial t}+\frac{d \mathbf{x}}{d t}\cdot\frac{\partial f_s}{\partial\mathbf{x}}+\frac{d \mathbf{v}}{d t}\cdot\frac{\partial f_s}{\partial \mathbf{v}}=0.\label{eq:vlasov}
\end{equation}
Hence, the Vlasov equation states that the distribution function is conserved along the particle trajectories in phase-space, and we see that these trajectories are the characteristics of the PDE.  

One can construct `macroscopic' quantities in real-space by taking velocity-space moments of the DF. For example, the number density (a scalar), bulk flow (a vector), and pressure tensor (of rank-2) for species $s$ are given by
\begin{eqnarray}
n_s(\mathbf{x};t)&=&\int\, f_s\, d^3v,\nonumber\\
\mathbf{V}_s(\mathbf{x};t)&=&n_s^{-1}\int \, \mathbf{v}\,f_s\,d^3v,\nonumber\\
P_{ij}(\mathbf{x};t)&=&\sum_sm_s\int (v_i-V_{i,s})(v_j-V_{j,s})\, f_s\, d^3v,\label{eq:pij}
\end{eqnarray}
for $m_s$ the mass of particle species $s$, and $\int d^3v$ the integral over all velocity space, i.e. $\mathbb{R}^3$. In order for the DF to be have physical meaning, one must be able to take velocity moments of any order, and the DF must be non-negative over all phase space (e.g. see \citet{Schindlerbook}),
\begin{eqnarray}
\bigg| \, \int v_x^i\,v_y^j\,v_z^k\, f_s\, d^3v\,\bigg| \,&<&\,\infty \, \, \forall \, i, \, j,\, k \in 0,1,2,... \label{eq:moments}\\
f_s\left(  \mathbf{x},\mathbf{v},t  \right) &\ge & 0 \, \forall \, \left(\mathbf{x},\mathbf{v},t  \right). \label{eq:nonneg}
\end{eqnarray}

\subsection{Vlasov-Maxwell equilibria}
This paper will focus on the application to a fully ionised and non-relativistic ion-electron plasma, for which the electromagnetic forces dominate the collective behaviour of the plasma. We shall also ignore gravitational effects. In such circumstances, an individual particle of charge $q_s$ and mass $m_s$ is subject to the Lorentz force,
\begin{equation}
m_s\frac{d\mathbf{v}}{dt}=q_s(\mathbf{E}+\mathbf{v}\times\mathbf{B}),\label{eq:lorentz}
\end{equation}
for $\mathbf{E}$ and $\mathbf{B}$ the self-consistent electric and magnetic fields respectively. A Vlasov equilibrium corresponds to a statistically steady-state of the particle distribution, and is mathematically described by $\partial_t f_s=0$. Therefore, we see from equations (\ref{eq:vlasov}) and (\ref{eq:lorentz}) that a collisionless plasma equilibrium DFs obeys by the steady-state \emph{Vlasov-Maxwell} equation,
\begin{equation}
\mathbf{v}\cdot\frac{\partial f_s}{\partial \mathbf{x}}+\frac{q_s}{m_s}(\mathbf{E}+\mathbf{v}\times\mathbf{B})\cdot\frac{\partial f_s}{\partial\mathbf{v}}=0.\label{eq:ssvm}
\end{equation}

Most typically (e.g. see \citet{Schindlerbook}), analytical equilibrium solutions of the Vlasov equation are constructed using a theorem attributed to Jeans \citep{Jeans-1915, Lynden-Bell-1962}. `Jeans' theorem' states that $f_s$ is a solution of the Vlasov equation if it is a function of a subset of the $k$ known constants of motion for particle species $s$,
\[
\{ C_{n,s}(\mathbf{x}(t),\mathbf{v}(t)) : d_tC_{n,s}=0, n=1,...,k \}. 
\]
Since $d_tC_{n,s}=\partial_t C_{n,s}=0$, we see that any such function will \emph{also be} a Vlasov equilibrium ($\partial_t f_s=0$). We should mention another (numerical) method to solve the Vlasov equation, that exploits the fact that whilst the DF must be a single-valued function over phase-space, $(\mathbf{x},\mathbf{v})$, it need not necessarily be a single-valued as a function of the constants of motion, $C_{n,s}(\mathbf{x},\mathbf{v})$, since $C_{n,s}$ may not be monotonic functions of $(\mathbf{x},\mathbf{v})$ (e.g. see \citet{Belmont-2012} and references therein).

Whilst the steady-state Vlasov-Maxwell equation (equation \ref{eq:ssvm}) can trivially be solved by any function of the constants of motion (that also obeys equations (\ref{eq:moments}) and (\ref{eq:nonneg})), the challenge lies in the fact that the DF must also be self-consistent with the time-independent Maxwell equations,
\begin{eqnarray}
\nabla\cdot\boldsymbol{E}&=&\frac{\sigma}{\epsilon_0},\label{eq:Gauss}\\
\nabla\times\boldsymbol{E}&=&\mathbf{0},\label{eq:Faraday}\\
\nabla\cdot\boldsymbol{B}&=&0,\label{eq:Solenoid}\\
\nabla\times\boldsymbol{B}&=&\mu_{0}\boldsymbol{j},\label{eq:Ampere}
\end{eqnarray}
for the speed of light $c=(\mu_0\epsilon_0)^{-1/2}$, $\epsilon_0$ the vacuum permittivity, $\mu_0$ the vacuum permeability, $\sigma$ the charge density, and $\mathbf{j}$ the current density. Maxwell's equations couple to the Vlasov-Maxwell equation through the source terms on the right-hand side (RHS) of Gauss' law, and Amp\`{e}re's Law (equations (\ref{eq:Gauss}) and (\ref{eq:Ampere}) respectively):
\begin{eqnarray}
\sigma (\mathbf{x})=\sum_sq_sn_s=\sum_sq_s\int f_s \, d^3v,\label{eq:sigma}\\
\mathbf{j} (\mathbf{x})=\sum q_sn_s\mathbf{V}_s=\sum_sq_s\int\mathbf{v}f_s \, d^3v.\label{eq:j}
\end{eqnarray}
Note that equations (\ref{eq:Faraday}) and (\ref{eq:Solenoid}) are automatically satisfied for the electric field $\mathbf{E}=-\nabla\phi$ defined as the (negative) gradient of the scalar potential $\phi(\mathbf{x})$, and $\mathbf{B}=\nabla\times\mathbf{A}$ the curl of the vector potential $\mathbf{A}(\mathbf{x})$.

An equilibrium solution of the \emph{Vlasov-Maxwell system} is therefore characterised by a self-consistent set of DFs and potential functions, 
\[\{f_s : s=i,e\},\mathbf{\phi}(\mathbf{x}),\mathbf{A}(\mathbf{x}),\] 
such that equations (\ref{eq:ssvm}), (\ref{eq:Gauss}) and (\ref{eq:Ampere}) are satisfied, with the subscripts $i$ and $e$ corresponding to ions and electrons respectively (e.g. see \citet{Mynick-1979a, Greene-1993, Schindlerbook}).

\subsection{The forward and inverse approaches}
The Vlasov-Maxwell system is an integro-differential set of equations, and there are two different approaches that could be made to solve it. The one most typically seen is the `forward' approach, in which one specifies the DFs, proceeds to calculate the source terms via the integrals in equations (\ref{eq:sigma}) and (\ref{eq:j}), and then attempts to solve the differential equations (\ref{eq:Gauss}) and (\ref{eq:Ampere}) for $\phi$ and $\mathbf{A}$ (e.g. see discussions in \cite{Bennett-1934, Grad-1961, Harris-1962, Sestero-1964, Sestero-1965, Lee-1979JGR, Schindlerbook, Kocharovsky-2010, Vasko-2013}). In this approach (`micro'$\to$`macro'), one typically assumes a form of the equilibrium DF that is thought to be mathematically/physically reasonable, and frequently one of Maxwellian form,
\[
f_{\text{Maxw},s} = \frac{n_s(\mathbf{x})}{(\sqrt{2\pi}v_{\text{th},s})^3}e^{-(\mathbf{v}-\mathbf{V}_s(\mathbf{x}))^2/(2v_{\text{th},s}^2)},
\]
for $v_{\text{th},s}$ a positive constant (the \emph{thermal velocity}). Then, given some conditions on certain elements of the macroscopic description (e.g. boundary conditions for the electromagnetic potentials), one is hopefully able to solve differential equations and calculate the macroscopic quantities $\mathbf{E}$ and $\mathbf{B}$, as well as $n_s, \mathbf{V}_s, \mathbf{j}, P_{ij}$ etc (e.g. see \citet{Harris-1962, Schindlerbook,Vasko-2013}). Whilst this approach is typically easier than the one that we are about to introduce, it has one significant drawback. In contrast to collisional plasmas for which there is in principle a unique equilibrium solution of the corresponding kinetic equation: the Maxwellian DF \citep{Grad-1949b}, collisionless plasmas admit an infinity of possible equilibrium DFs (i.e. any function of the constants of motion). Hence, for the many man-made, terrestrial, space, solar and astrophysical plasmas in which the collisionality is insufficient to drive the DF to a thermal equilibrium Maxwellian DF, one does not \emph{a priori} know what form of the DF on which to base their calculation. 

The `inverse' approach is the method that we shall focus on in this paper (`macro'$\to$`micro'). In the `inverse problem for collisionless equilibria' (IPCE), one specifies certain features of the macroscopic equilibrium configuration, e.g. $\mathbf{B}$ and $\mathbf{E}$, and attempts to find a self-consistent DF (e.g. see discussions in  \cite{Alpers-1969, Channell-1976, Mynick-1979a,Greene-1993, Harrison-2009POP, Belmont-2012, Allanson-2016JPP}). This approach involves the inversion of integral equations, and has two potential drawbacks. The first is that there are in principle infinitely many solutions to IPCE, and so one has to question whether the DF obtained by the chosen method is physically realistic, or just a mathematical curiosity. The second is that the equilibrium DF may not be able to be expressed in closed form, and the necessary conditions of equations (\ref{eq:moments}) and (\ref{eq:nonneg}) may be in question. 

\subsection{Previous work on non-negativity}
The first body of works on IPCE \citep{Bernstein-1957, Montgomery-1969, Tasso-1969, Schamel-1971} mainly focussed on electrostatic configurations, i.e. $\mathbf{E}\ne\mathbf{0}$ and $\mathbf{B}=\mathbf{0}$ (the \emph{Vlasov-Poisson} system). In \citet{Bernstein-1957}, an inductive integral equation method is developed that calculates the DF of trapped electrons in a nonlinear travelling electrostatic wave (\emph{Bernstein-Greene-Kruskal (BGK) waves}), for a given 1D scalar potential, $\phi$, in the wave frame. Whilst \citet{Bernstein-1957} recognised that equation (\ref{eq:nonneg}) must be satisfied for the DF to be physically meaningful, they did not formally explore this feature. \citet{Montgomery-1969}, \citet{Tasso-1969} and \citet{Schamel-1971} demonstrated that it is indeed possible to obtain non-negative trapped DFs for the trapped electron population, and this is relevant to the physics of nonlinear plasma waves (e.g. see \citet{Schamel-1972JPP,Schamel-1986}) and collisionless shocks (e.g. see \citet{Burgessbook}).

However, the work in this paper will focus on IPCE applied to `strictly neutral' magnetised plasma configurations $(\phi=|\mathbf{E}|=\sigma=0$ and $\mathbf{B}\ne\mathbf{0})$, as used in Vlasov-Maxwell equilibrium studies by \citet{Grad-1961, Hurley-1963, Nicholson-1963, Schmid-Burgk-1965, Moratz-1966, Lerche-1967, Alpers-1969, Channell-1976, Bobrova-1979, Lakhina-1983, Attico-1999, Bobrova-2001, Fu-2005, Yoon-2005, Harrison-2009PRL, Neukirch-2009, Panov-2011, Wilson-2011, Belmont-2012, Janaki-2012, Abraham-Shrauner-2013, Ghosh-2014, Kolotkov-2015,Allanson-2015POP,Allanson-2016JPP}. This approach is reasonable when typical spatial variations, $L$, are much larger than a quantity known as the \emph{Debye radius}, $\lambda_D$,
\begin{equation}
\frac{\lambda_D}{L}\ll 1,\hspace{3mm}\text{s.t.}\hspace{3mm}\lambda_D=\sqrt{\frac{\epsilon_0k_{B}T_e}{n_ee^2}},\nonumber
\end{equation}
for $k_{B}$ Boltzmann's constant, $T_e$ the electron temperature, and $e$ the fundamental charge \citep{Schindlerbook}.

Questions regarding the non-negativity of IPCE solutions for magnetised plasmas arise in the works by \citet{Abraham-Shrauner-1968, Alpers-1969, Channell-1976, Hewett-1976, Suzuki-2008,Allanson-2015POP,Allanson-2016JPP}. All of these works, except \citet{Suzuki-2008}, consider equilibria that are 1D in space, and use (possibly infinite) expansions in Hermite polynomials \citep{Arfkenbook} to represent the DF, i.e. a non-closed form. In the following section, we proceed to develop the basic formalism required for IPCE with 1D neutral Vlasov-Maxwell equilibria, using the constants of motion approach.

\section{One-dimensional strictly neutral Vlasov-Maxwell equilibria}

Without loss of generality, we take the $z$ coordinate as the one spatial coordinate on which the 1D system dynamics depend. As a result of this assumption, the magnetic field and current density are written as 
\begin{eqnarray}
\mathbf{B}=\left(-\frac{dA_y}{dz},\frac{dA_x}{dz},0\right),\nonumber\\
\mathbf{j}=\frac{1}{\mu_0}\left(-\frac{dB_y}{dz},\frac{dB_x}{dz},0\right),\nonumber
\end{eqnarray}
respectively. Furthermore, the classical action (for a particle of species $s$),
\[
S=\int_{t_1}^{t_2} L \, dt=\int_{t_1}^{t_2}(m_sv^2/2+q_s\mathbf{v}\cdot\mathbf{A})\, dt,
\]
for $t_1,t_2,t\in\mathbb{R}$, is invariant ($\delta S=0$) under infinitesimal continuous transformations in $t,x$ and $y$ (e.g. see \citet{Landaufields}). Since the system is invariant in both time and two spatial dimensions, we have -- by Noether's theorem (e.g. see \citet{Weinbergbook}) -- three known constants of motion for a particle in an electromagnetic field, 
\begin{eqnarray}
H_s&=&m_sv^2/2,\nonumber\\
p_{x,s}&=&m_sv_x+q_sA_x,\nonumber\\
p_{y,s}&=&m_sv_y+q_sA_y,\nonumber,
\end{eqnarray} 
the Hamiltonian, and canonical momenta in the $x$ and $y$ directions respectively. We shall now make one broad assumption on the functional form of the DF, namely that -- as a function of the constants of motion -- it is written as,
\begin{equation}
f_s=f_s(H_s,p_{x,s},p_{y,s})=\frac{n_{0}}{(\sqrt{2\pi v_{\text{th},s}})^3}e^{-\beta_sH_s}g_s(p_{x,s},p_{y,s}),\label{eq:ansatz}
\end{equation}
for $n_{0},v_{\text{th},s}$, and $\beta_s=(m_sv_{\text{th},s}^2)^{-1}$ positive constants/parameters, with dimensions of number density, velocity and $1/$energy respectively; and $g_s$ an unknown function to be determined. This assumption has a long history in the literature (e.g. see first use in \citet{Alpers-1969,Channell-1976}), and is chosen for both mathematical reasons (integrability), and physical ones (the DF reduces to a `stationary' Maxwellian when $g_s=1$). Hence, the task of IPCE has been reduced to finding $g_s$ functions that are self-consistent with the prescribed macroscopic conditions. As written, it is the $g_s$ function that potentially encodes the interesting non-Maxwellian phase-space structure, only permitted for collisionless plasma equilibria.

One immediate consequence of the ansatz in equation (\ref{eq:ansatz}) is that $V_{z,s}=0$, since the DF is an even function of $v_z$. As such, the $zz$ component of the pressure tensor (equation \ref{eq:pij}) is written as
\begin{equation}
P_{zz}=\sum_sm_s\int v_z^2f_s d^3v.\label{eq:pzz}
\end{equation}
Many authors have noted the pivotal role that $P_{zz}$ plays in the Vlasov-Maxwell equilibrium system for 1D plasmas (e.g. see \citet{Grad-1961, Channell-1976, Mynick-1979a, Greene-1993, Tassi-2008,Harrison-2009POP}), and the following discussion shall make use of results from these works. 

The $zz$ component of $P_{ij}$ is not the only non-zero component for our problem, but it is the only one that plays a role in macroscopic force balance, given for a neutral plasma as
\begin{eqnarray}
\sum_{i=1}^3\frac{\partial}{\partial x_i}P_{ij}=(\mathbf{j}\times\mathbf{B})_j,\nonumber\\
\implies \frac{d}{dz}P_{zz}=j_xB_y-j_yB_x=-\frac{1}{2\mu_0}\frac{d}{dz}B^2\label{eq:pressurebalance}.
\end{eqnarray} 
Equation (\ref{eq:pressurebalance}) is the statement of pressure balance, $P_{zz}+B^2/(2\mu_0)=P_{\text{total}}$, for $B^2/(2\mu_0)$ the magnetic energy density (pressure), and $P_{\text{total}}$ the total pressure (thermal plus magnetic). Furthermore, the existence of a neutral Vlasov-Maxwell equilibrium can be shown to imply that $P_{zz}=P_{zz}(A_x(z),A_y(z))$ \citep{Channell-1976}, and hence the chain rule gives
\begin{equation}
\frac{d}{dz}P_{zz}=\frac{\partial P_{zz}}{\partial A_x}\frac{dA_x}{dz}+\frac{\partial P_{zz}}{\partial A_y}\frac{dA_y}{dz}\label{eq:chainrule}.
\end{equation}
By comparing terms in the first equality of equation (\ref{eq:pressurebalance}), and equation (\ref{eq:chainrule}), we see that $P_{zz}$ plays the role of a potential function in this problem, from which the sources of Amp\`{e}re's Law (equation (\ref{eq:Ampere})) can be derived
\begin{eqnarray}
-\mu_0\frac{d^2A_x}{dz^2}=j_x(z)=\frac{\partial P_{zz}}{\partial A_x},\label{eq:Amp1}\\
-\mu_0\frac{d^2A_y}{dz^2}=j_y(z)=\frac{\partial P_{zz}}{\partial A_y}.\label{eq:Amp2}
\end{eqnarray}
These equations are analogous to the equations of motion for a particle at `position' $(A_x,A_y)$ and `time' $z$, under the influence of a `potential' $P_{zz}$. Equations (\ref{eq:pzz}), (\ref{eq:Amp1}) and (\ref{eq:Amp2}) neatly summarise the task for a self-consistent solution of the neutral Vlasov-Maxwell equilibrium system. 

\subsection{The inverse problem for one-dimensional Vlasov-Maxwell equilibria}\label{sec:1dvm}
In the context of IPCE, after one has first specified the macroscopic equilibrium, i.e. given $(A_x(z),A_y(z))$, the first step is to calculate a $P_{zz}$ function that satisfies equations (\ref{eq:Amp1}) and (\ref{eq:Amp2}). For example, in the case of `force-free' magnetic fields \citep{Marshbook}, there is an algorithmic path that takes $(A_x(z),A_y(z))$ as input, and gives $P_{zz}(A_x,A_y)$ as output (e.g. see \citet{Harrison-2009PRL, Neukirch-2009, Wilson-2011, Abraham-Shrauner-2013, Kolotkov-2015, Allanson-2015POP, Allanson-2016JPP}). 

The next task is to invert equation (\ref{eq:pzz}) for a known left-hand side (LHS), and the unknown function $f_s$. The details of this step are summarised in \citet{Channell-1976}. Channell's method is characterised by the inversion of the following equation for $g_s$,
\begin{eqnarray}
P_{zz}(A_x,A_y)=\frac{\beta_{e}+\beta_{i}}{\beta_{e}\beta_{i}}\frac{n_{0}}{2\pi m_{s}^2v_{\text{th},s}^2}\int\int \; e^{-\beta_{s}\left((p_{x,s}-q_{s}A_x)^2+(p_{y,s}-q_{s}A_y)^2\right)/(2m_{s})}g_{s}(p_{x,s},p_{y,s})dp_{x,s}dp_{y,s}, \label{eq:Channell}
\end{eqnarray}
which is a re-expression of equation (\ref{eq:pzz}) (after one layer of integration over $v_z$, with the functional form of the DF given by the ansatz in equation (\ref{eq:ansatz}), and the integral taken over $\mathbb{R}^2$). Note that $P_{zz}$ is formally defined as a sum (over species) of integrals, whereas the RHS of equation (\ref{eq:Channell}) has only one integral, indexed by a generic species $s$, i.e. the RHS must yield the same result for $s=i$ as for $s=e$. This requirement is shown by Channell to be equivalent to that imposed by exact charge neutrality, 
\[
n_{i}(A_x,A_y)=n_e(A_x,A_y).
\]
After some consideration, we should be convinced that the species-independent result of the RHS of equation (\ref{eq:Channell}) implies that the $g_s$ function must itself depend on species-dependent parameters, $g_s=g_s(p_{x,s},p_{y,s};\varepsilon_s)$. Using this fact, and after making some substitutions, equation (\ref{eq:Channell}) can be re-written according to
\begin{eqnarray}
\mathcal{P}_s(\mathcal{A}_{x,s},\mathcal{A}_{y,s})=\frac{1}{4\pi\varepsilon_s}\int\int \; e^{-\left((p_{x,s}-\mathcal{A}_{x,s})^2+(p_{y,s}-\mathcal{A}_{y,s})^2\right)/(4\varepsilon_s)}g_{s}(p_{x,s},p_{y,s};\varepsilon_s)dp_{x,s}dp_{y,s}, \label{eq:normChannell}
\end{eqnarray}
with $\varepsilon_s=m_s^2v_{\text{th},s}^2/2$, $\boldsymbol{\mathcal{A}}_s=q_s\mathbf{A}$, and with $\mathcal{P}_s$ defined according to
\begin{equation}
\mathcal{P}_s(\mathcal{A}_{x,s},\mathcal{A}_{y,s})=\frac{\beta_e\beta_i}{(\beta_e+\beta_i)n_{0}}P_{zz}\left(A_x,A_y\right).
\end{equation}

\subsection{The Weierstrass transform}
Equations (\ref{eq:Channell}) and  (\ref{eq:normChannell}) express $P_{zz}$ and $\mathcal{P}_{s}$ as two-dimensional (2D) integral transforms of the $g_s$ function. As discussed in \citet{Allanson-2015POP,Allanson-2016JPP}, the integral transform in equation (\ref{eq:normChannell}) is a 2D generalisation of the Weierstrass transform \citep{Widder-1951,Widder-1954,Bilodeau-1962, zayedbook}. The Weierstrass transform, $u(x,t)$, of $u_0(y)$ is defined by
\begin{equation}
u(x,t):=\mathcal{W}\left[u_0 \right] (x,t)=\frac{1}{\sqrt{4\pi t}}\int \, e^{-(x-y)^2/(4t)}\,u_0(y)\,dy,\label{eq:Weierstrass}
\end{equation}
for $x,y,t\in\mathbb{R}$ and $t> 0$. This is also known as the Gauss transform, Gauss-Weiertrass transform or the Hille transform \citep{Widder-1951}. As the Green's function solution to the 1D heat/diffusion equation on an infinite domain,
\[
\frac{\partial u}{\partial t}-\frac{\partial^2 u}{\partial x^2} =0\nonumber,
\]
with initial data $u_0(x)$, $u(x,t)$ represents the temperature/density profile on an infinite rod, $t$ seconds after it was $u_0(x)$ (e.g. see \citet{Widder-1951}). Note that equation (\ref{eq:Weierstrass}) is only a meaningful representation of the function $u(x,t)$ (i.e. the Weierstrass transform exists) if $u_0(x)$ is locally integrable, and such that
\[
|u_0(x)|\le Me^{\alpha x^2},
\]
for $M<\infty$ and $0<t<1/(4\alpha)$ (e.g. see \citet{Widder-1951,zayedbook}). One can immediately see from equation (\ref{eq:Weierstrass}) that the Weierstrass transform of an everywhere non-negative function is itself a non-negative function, and furthermore that the transform of an everywhere negative function is an everywhere negative function. However, it is possible for the Weierstrass transform of a function that is somewhere negative (i.e. a candidate $g_s$ function), to be a function that is everywhere non-negative (i.e. the $P_{zz}$ function). This case is potentially very troubling for a meaningful solution of IPCE, and is discussed in \citet{Allanson-2016JPP}, and later on in this paper. 

Motivated by the appearance of the Weierstrass transform in IPCE, and for completeness, we now introduce some basic properties of the initial value problem (IVP) for the heat equation.

\subsection{Initial value problems for the heat equation}
The $n$-dimensional ($n$D) heat equation models the temperature distribution, $u(x,t)$, on an infinite $n$D spatial domain, and is given by 
\[
\frac{\partial u}{\partial t}-\nabla^2u =0\nonumber,
\]
for $\nabla^2=\sum_{i=1}^n\partial^2_{x_i}$. We define the IVP for the $n$D heat equation on the unbounded spatial domain $\mathbb{R}^n$, according to
\begin{eqnarray}
u=u(x,t)\in C^2(\mathbb{R}^n\times\Omega_T)\cap C^0(\mathbb{R}^n\times \overline{\Omega_T}),\nonumber\\
\frac{\partial u}{\partial t}-\nabla^2u =0\hspace{3mm}\text{in}\hspace{3mm}\mathbb{R}^n\times \Omega_T\nonumber,\\
u(x,0)=u_0(x),\hspace{3mm}x\in\mathbb{R}^n,\nonumber\\
\sup_{(x,t)\in\mathbb{R}^n\times\Omega_T}|u(x,t)|= u_{\text{B}}(x)\nonumber ,
\end{eqnarray}
for the as yet unspecified temporal domain $\Omega_T$, for $\overline{\Omega_T}$ the closure of $\Omega_T$, and for $u_{\text{B}}(x)$ the as yet unspecified supremum of $u(x,t)$. There are some standard results regarding bounded, unbounded, and non-negative solutions of the IVP respectively, and we shall briefly detail these.

There is a unique bounded solution of the IVP for bounded and continuous initial data, $u_0$, i.e. $\Omega_T=(0,\infty)$ and $u_{\text{B}}(x)=C<\infty$ (e.g. see \citet{sauvignybook}). This unique solution is defined by the $n$D integral transform (a generalisation of the Weierstrass transform) given by
\begin{equation}
u(x,t)=(4\pi t)^{-n/2}\int \, e^{-(x-y)^2/(4t)}\,u_0(y)\,dy,\label{eq:Heat_inverse}
\end{equation}
and belongs to $C^\infty$ for $x\in\mathbb{R}^n$, $t>0$. Moreover, $u$ has the initial values $u_0$, in the sense that when we extend $u$ by $u(x,0)=u_0(x)$ to $t=0$, then $u$ is continuous for $x\in\mathbb{R}^n$, $t\ge 0$.

It is also possible to obtain unbounded solutions of the heat equation, using equation (\ref{eq:Heat_inverse}). In fact, there is a unique solution to the IVP on the bounded temporal domain $\Omega_T=(0,T)$, with $u_{\text{B}}(x)=M e^{\alpha x^2}$, and such that $4\alpha T<1$ (e.g. see \citet{johnbook}).

Of clear interest to the work in this paper, are solutions to the IVP for non-negative functions. It is a standard result that there is a unique solution to the IVP on $\Omega_T=(0,T)$ using equation (\ref{eq:Heat_inverse}), with the condition that $u(x,t)$ is non-negative on $\mathbb{R}^n\times [0,T)$ (e.g. see \citet{johnbook}).

In informal terms, we see that a non-negative initial condition (or history) for the heat distribution implies a non-negative distribution in the present and in the future. However, the converse is not necessarily true. The extent to which we can be sure of a non-negative `past', given a non-negative `present', is the question we consider. Tackling this inverse problem, in the context of equation (\ref{eq:normChannell}), is perhaps the main mathematical challenge for validity of solutions obtained in IPCE, and is akin to going backwards in time (see \citet{Evansbook} for a brief discussion on `backwards solutions of the heat equation'). Our known and non-negative `present distribution' is defined by $\mathcal{P}_s(\mathcal{A}_{x,s},\mathcal{A}_{y,s})$, and the `past distribution', with questionable sign, is defined by $g_s(p_{x,s},p_{y,s};\varepsilon_s)$.

\subsection{Implications and interpretation for IPCE}
Equation (\ref{eq:normChannell}) casts the inverse problem in direct comparison with the Weiertrass transform, thus making a correspondence between space and time in the heat equation, $(x, t)$, to $(\mathbf{A},\varepsilon_s)$ in our inverse problem. However, one difference is that the $g_s$ function must - at least parametrically - depend on `time', $\varepsilon_s$, in contrast to the initial condition (i.e. a time-independent function) that is part of the integrand in Equation (\ref{eq:Weierstrass}). We know that $g_s$ must depend on $\varepsilon_s$, since the result of the integral (the LHS) must be independent of $\varepsilon_s$, as mentioned in Section \ref{sec:1dvm}. Hence, it is not immediately clear how `far' the analogy applies, e.g. is there a differential equation (the heat equation or similar) that $g_s$ and/or $P_{zz}$ obey?  

On this subject, and in contrast to the IVP, the approach of IPCE casts the $P_{zz}$ function as the given/fixed quantity, i.e. the \emph{final condition}. In tackling IPCE, we look for non-negative \emph{`initial conditions'}, $g_{s}$, that will produce the correct $P_{zz}$. Hence, it is reasonable that the `initial condition' should be `time-dependent'. That is to say, \emph{``given a $P_{zz}(A_x,A_y)$ function, we calculate the self-consistent $g_s$ function `$\varepsilon_s$ seconds ago' by inverting equation (\ref{eq:normChannell}), the integral transform that `evolves' the $g_s(p_{x,s},p_{y,s};\varepsilon_s)$ function by `$\varepsilon_s$ seconds' ''.}

  There is one more conceptual hurdle to overcome, that has it's origins in a physical problem rather than a mathematical one. Properly considered, the LHS of equation (\ref{eq:Channell}) is a function of the (macroscopic) vector potential, which is typically normalised using the macroscopic parameters $B_0$ and $L$ ($\tilde{\mathbf{A}}=\mathbf{A}/(B_0L)$): constants with dimensions of magnetic field and length respectively. In contrast, the integral on the RHS is in (microscopic) momentum space, which is typically normalised using the microscopic parameters $m_s$ and $v_{\text{th},s}$ ($\tilde{\mathbf{p}}_s=\mathbf{p}_s/(m_sv_{\text{th},s})=\mathbf{p}_s/(\sqrt{2\varepsilon_s})$). Therefore, one has to fix a relationship between the micro- and macroscopic parameters (e.g. see \citet{Neukirch-2009, Wilson-2011, Kolotkov-2015, Allanson-2015POP} for practical examples). This can be achieved by tuning the `time' using a new parameter, $\delta_s$, according to
  \begin{equation}
\left(\frac{m_s^2v_{\text{th},s}^2}{2}=:\right)\varepsilon_s:=\frac{(eB_0L)^2\delta_s^2}{2}, \label{eq:tuning}
\end{equation}
for $e$ the fundamental charge, and $\delta_s$ the dimensionless and species-dependent \emph{magnetisation parameter} (e.g. see \cite{Fitzpatrickbook}), defined by 
\[
\delta_s=\frac{r_{L,s}}{L}=\frac{m_{s}v_{\text{th},s}}{eB_0L}.
\]
It is the ratio of the thermal Larmor radius, $r_{Ls}=v_{\text{th},s}/|\Omega_s|$, to the characteristic length scale of the system, $L$. The gyrofrequency of particle species $s$ is $\Omega_s=q_{s}B_0/m_{s}$. As formulated here, the magnetisation parameter is the dimensionless parameter linking the microscopic and macroscopic descriptions of the Vlasov-Maxwell equiibrium.

\subsection{Using Fourier transforms to solve IPCE}\label{sec:Ftransform}
As written, Equation (\ref{eq:normChannell}) defines $\mathcal{P}_s(\mathcal{A}_{x,s},\mathcal{A}_{y,s})$ as a 2D \emph{convolution} of the functions $G_{s}=e^{-(p_{x,s}^2+p_{y,s}^2)/4\varepsilon_s}$ and $g_s(p_{x,s},p_{y,s})$, with the convolution of functions $h_1(\mathbf{x})$ and $h_2(\mathbf{x})$ defined as
\begin{eqnarray}
h_1\star h_2\, (\mathbf{x}_1)&=&\int h_1(\mathbf{x}_1-\mathbf{x})h_2(\mathbf{x})d\mathbf{x}\nonumber =\int h_1(\mathbf{x})h_2(\mathbf{x}_1-\mathbf{x})d\mathbf{x},\label{eq:cconvolution}
\end{eqnarray}
with $\mathbf{x},\mathbf{x}_1\in\mathbb{R}^n$ and the integrals over $\mathbb{R}^n$.  The $n$D Fourier transform (FT) of a function is defined by
\[
\text{FT}[h(\mathbf{x})]:=\tilde{h}(\mathbf{k})= (2\pi)^{-n/2} \int e^{-i \mathbf{k}\cdot\mathbf{x}} h(\mathbf{x}) d\mathbf{x},
\]
for $\mathbf{k}\in\mathbb{R}^n$, and the integral over $\mathbb{R}^n$. There is a very useful result regarding the FT of a convolution. The convolution theorem states that
\[
\text{FT} [h_1\star h_2] = \tilde{h}_1 \tilde{h}_2 ,
\]
\citep{zayedbook}. That is to say that the Fourier transform of a convolution of functions is the product of the transforms of the individual functions. As such, and using the convolution theorem, $g_s$ can - at least formally - be written
\begin{equation} 
g_s(p_{xs},p_{ys};\varepsilon_s)=4\pi\varepsilon_s\text{IFT}\left[    \, \tilde{P}_{zz} \,/ \,\tilde{G_{s}}  \,            \right],\label{eq:formalfourier}
\end{equation}
for IFT the $n$D inverse Fourier transform,
\[
\text{IFT}[\tilde{h}(\mathbf{k})]=h(\mathbf{x})=(2\pi)^{-n/2}\int  e^{i \mathbf{k}\cdot\mathbf{x}} \tilde{h}(\mathbf{k}) d\mathbf{k} .
\]
This Fourier transform method has been used by authors to solve IPCE (e.g. see \citet{Channell-1976, Harrison-2009PRL}. Indeed, at first glance, it would seem that using equation (\ref{eq:formalfourier}) is the general solution to our IPCE problem. However, the solution defined is only formal, without further investigation. It is not of use when either the Fourier transform of the $g_s$ or $\mathcal{P}_s$ functions cannot be evaluated, or when the inverse Fourier transform expression on the RHS of equation (\ref{eq:formalfourier}) cannot be evaluated. It may be the case that these transforms cannot be evaluated either in the sense that there exists no exact analytic answer in closed form, or that they are divergent (e.g. see \citet{Channell-1976, Allanson-2016JPP} for examples and discussion). 

The subsequent work in this paper makes use of, and develops, the theory of solutions to IPCE with the use of Hermite polynomial expansions. This technique is to be seen as an alternative method to Fourier transforms.

\subsection{Summary}
In this section we have demonstrated that for neutral equilibria, IPCE in 1D can be reduced to the inversion of Weierstrass transforms once the expression for the pressure tensor component $P_{zz}$ is known. The problem is therefore succinctly described by equations (\ref{eq:Amp1}), (\ref{eq:Amp2}) and (\ref{eq:normChannell}). We have discussed the parallels between the integral equations to be inverted, and the Green's function solutions of the heat equation. Whilst there are many standard results regarding the IVP for the heat equation, the problem that we face with IPCE is more related to `backwards solutions': \emph{`for a given distribution that is everywhere non-negative heat distribution at time $t_1$, was there a everywhere non-negative distribution at time $t_0$, in the past?'}.

\section{Expansions in Hermite polynomials}
\subsection{Hermite polynomials}
The use of Hermite polynomials in kinetic theory dates back, at least, to \citet{Grad-1949b} in the study of rarefied collisional gases, in which non-equilibrium DFs are represented by shifted Maxwellians multiplied by an expansion in ``$n$-dimensional'' Hermite polynomials \citep{Grad-1949a}. However, the most typical approach in collisionless and weakly collisional plasma kinetic theory is to use expansions in `scalar' Hermite polynomials \citep{zayedbook}, defined by
\begin{eqnarray}
H_n(p)&=&(-1)^ne^{p^2}\frac{d^{n}}{dp^{n}}e^{-p^2},\label{eq:hdef}\\
\int_{-\infty}^\infty H_m(p)H_n(p)e^{-p^2}dp&=&\delta_{m,n}2^nn!\sqrt{\pi},\label{eq:orthogonal}
\end{eqnarray}
for $\delta_{m,n}$ the Kronecker delta, and $p\in\mathbb{R}$. Hermite polynomials are a complete orthogonal set of polynomials for $g(p)\in L^2(\mathbb{R},e^{-p^2}dp)$ \citep{Sansonebook, Arfkenbook}. That is to say that for any piecewise continuous $g(p)$, such that
\begin{equation}
\int_{-\infty}^\infty |g(p)|^2 e^{-p^2}dp <\infty ,\label{eq:squareintegrable}
\end{equation}
then there exists an (infinite) expansion in Hermite polynomials, $\sum_{n=0}^{\infty} c_n H_n(p)$, such that
\begin{equation}
\lim_{k\to\infty}\int_{-\infty}^\infty\bigg|g(p)-\sum_{n=0}^k c_n H_n(p)\bigg|^2e^{-p^2}dp=0.\label{eq:parity}
\end{equation}
Hermite polynomials have a long history in kinetic theory precisely due to equation (\ref{eq:orthogonal}): they are a natural orthogonal basis with which to use when also considering Gaussian/Maxwellian/normal profiles $\sim e^{-\tilde{p}^2} \sim e^{-\tilde{v}^2}$, for some appropriately normalised momenta or velocity ($\tilde{p}$ or $\tilde{v}$).

\subsection{Hermite polynomials for exact VM equilibria}
In the work by \citet{Abraham-Shrauner-1968}, expansions in Hermite polynomials of the canonical momentum are used to solve the VM system for the case of `stationary waves' in a manner similar to that to be described in this section. These wave structures correspond not to Vlasov equilibria, but rather to nonlinear waves that are stationary in the wave frame. Abraham-Shrauner considers a 1D plasma with only one component of current density, first in a general sense, and then considers three different magnetic field configurations. \citet{Alpers-1969} also presents a somewhat general discussion on the use of Hermite polynomials for 1D VM equilibria, and proceeds to consider models suitable for the magnetopause, with both one component of the current density, and with two. In the work by \citet{Channell-1976}, two methods are presented for the solution of the inverse problem with neutral VM equilibria, by means of example. These two methods are inversion by Fourier transforms and -- once again -- expansion over Hermite polynomials respectively. Channell uses Hermite polynomials in the canonical momenta, but this time with two components of the current density, for the specific case of a magnetic field that is especially suitable to be considered as a stationary wave solution.

In contrast to \citet{Abraham-Shrauner-1968, Alpers-1969, Channell-1976}, the works by \citet{Hewett-1976,Suzuki-2008} both consider the forwards problem in VM equilibria, and use Hermite polynomial expansions in velocity space, for 1D and 2D plasmas respectively. \citet{Hewett-1976} assume a representation for the DF using expansions in velocity space, but with only one current density component, and ensure self-consistency with Maxwell's equations numerically, whereas \citet{Suzuki-2008} use an analytical approach, e.g. demonstrating that the Hermite polynomial approach can reproduce known equilibria such as the Harris sheet \citep{Harris-1962}, and the Bennett Pinch \citep{Bennett-1934}. 

Crucially, none of the above references sytematically tackled the necessary mathematical conditions of convergence and non-negativity in a rigorous way. Motivated by new exact Vlasov-Maxwell solutions involving expansions in Hermite polynomials in \citet{Allanson-2015POP}, the work in \citet{Allanson-2016JPP} formally treated the use of Hermite polynomials in IPCE, and tackled the problems of convergence, boundedness and non-negativity of the resultant DF. This work will be discussed, and built upon, in the following sections.

To give a subset of (modern) examples outside the realm of equilibrium studies \emph{per se}, Hermite polynomial expansions are used by \citet{Daughton-1999} to assess the linear stability of a Harris current sheet; by \citet{Camporeale-2006} also on the linear stability problem, using a truncation method somewhat like that of \cite{Grad-1949b}, and managing to bypass the traditional approach of integrating over the `unperturbed orbits' \citep{Coppi-1966, Drake-1977, Quest-1981A, Daughton-1999}; by \citet{Zocco-2015} on linear collisionless Landau damping  \citep{Landau-1946,Mouhot-2011}; and by \citet{Schekochihin-2016} on the problem of the free-energy associated with velocity-space moments of the DF, in the problem of plasma turbulence.

\subsection{Formal solutions to 1D IPCE using Hermite polynomials}
Whereas Equations (\ref{eq:hdef}) and (\ref{eq:squareintegrable}) are the standard `physicists' definitions of Hermite polynomials, it will be of use in this work, as in \citet{Alpers-1969,Channell-1976,Allanson-2015POP,Allanson-2016JPP} to consider the scaled function $H_n(p/(2\sqrt{\varepsilon_s}))$. This slight modification results in changes to Equations (\ref{eq:hdef}), (\ref{eq:orthogonal}), (\ref{eq:squareintegrable}) and (\ref{eq:parity}), easily achieved by substitution.  

In fact, we see that expansions in Hermite polynomials $H_n(p/(2\sqrt{\varepsilon_s}))$ are a complete orthogonal set for $f\in L^2(\mathbb{R},e^{-p^2/(4\varepsilon_s)}dp)$. By equation (\ref{eq:squareintegrable}), we see that this means that expansions in Hermite polynomials, $H_n(p/(2\sqrt{\varepsilon_s}))$, are valid representations for piecewise-continuous functions $g$, such that $|g|\le Me^{p^2/(8\varepsilon_s)}$, for $M<\infty$. This condition is more strict than that for the existence of the integrals in equations (\ref{eq:normChannell}) and (\ref{eq:Weierstrass}), i.e. the validity of the Weierstrass transform representation (equivalent to $|g|\le Me^{p^2/(4\varepsilon_s)}$). 

Expansions in Hermite polynmials are of particular interest when the Fourier transform inversion technique detailed in Section (\ref{sec:Ftransform}) is intractable, and/or when $P_{zz}(A_x,A_y)$ is not given in closed form. In \citet{Allanson-2016JPP}, IPCE was solved using Hermite polynomial expansion, and the assumption was made that the Maclaurin expansion of $P_{zz}(A_x,A_y)$ was either `summatively' or `multiplicatively' separable in it's indices, i.e. of the form
\begin{equation}
P_{zz}\propto P_1(A_x)+P_2(A_y),\hspace{3mm}\text{or}\hspace{3mm}P_{zz}\propto P_{1}(A_x)P_2(A_y). \label{eq:p_allanson}
\end{equation}
Here we generalise the work presented in \citet{Allanson-2016JPP} to an arbitrarily indexed 2D sum, we suppose that $P_{zz}(A_x,A_y)$ is given as a 2D sum of the most general form,
\begin{equation}
P_{zz}(A_x,A_y)=n_{0}\frac{\beta_e+\beta_i}{\beta_e\beta_i}\sum_{m,n}c_{m,n}\left(\frac{A_x}{B_0L}\right)^m\left(\frac{A_y}{B_0L}\right)^n,\label{eq:pressureansatz}
\end{equation}
with the RHS a convergent Maclaurin expansion with infinite radius of convergence in both of it's arguments. The indices $m,n\in\{0,1,2,...\}$, and the coefficient $c_{m,n}\in\mathbb{R}$. (Note that convergence of the Maclaurin expansion of $P_{zz}$, and hence the convergence of the sum of derivatives, implies that $P_{zz}\in C^\infty (\mathbb{R}^2)$, e.g. see \citet{Bartle}).

Using theory as in \citet{Bilodeau-1962, Allanson-2016JPP}, it can be shown that the following expansion in Hermite polynomials 
\begin{equation}
g_s(p_{x,s},p_{y,s};\varepsilon_s)=\sum_{m,n}c_{m,n}\text{sgn}(q_s)^{m+n}\left(\frac{\delta_s}{\sqrt{2}}\right)^{m+n}H_m\left(\frac{p_{x,s}}{2\sqrt{\varepsilon_s}}\right)H_n\left(\frac{p_{y,s}}{2\sqrt{\varepsilon_s}}\right),\label{eq:gensol}
\end{equation}
is, formally speaking, an exact inverse solution of equation (\ref{eq:Channell}), for $P_{zz}$ given by equation (\ref{eq:pressureansatz}).

\subsection{Mathematical criteria}
Since a $g_s$ function found using the Hermite polynomial method - as in equation (\ref{eq:gensol}) - could be an infinite series of polynomials that does not represent a known function in closed form, it is by no means clear if $g_s$ is everywhere non-negative. This issue is recognised by \citet{Abraham-Shrauner-1968, Hewett-1976}. Not only is the non-negativity in question, but it is not obvious whether a given expansion in Hermite polynomials even converges, and this question was also raised by \citet{Hewett-1976}. Finally, even if the Hermite expansion converges, it must -when multiplied by the Maxwellian factor (equation (\ref{eq:ansatz})) - produce a DF for which velocity moments of all order exist, as discussed in Section 2. In order to have full confidence in the Hermite polynomial method we need to address these issues of non-negativity, convergence, and the existence of moments.

We should mention that the \emph{reverse} questions are well established, i.e. if one \emph{a priori} knows the DF in closed form, or at least if Equation (\ref{eq:squareintegrable}) is satisfied. In such circumstances, one can represent a given non-negative DF as a Maxwellian multiplied by an expansion in Hermite polynomials, $H_n(p/(2\sqrt{\varepsilon_s}))$, provided the $g_s$ function grows at a rate below $e^{p^2/(8\varepsilon_s)}\sim e^{v^2/4v_{\text{th},s}^2}$ \citep{Grad-1949b, Widder-1951}.

In \citet{Allanson-2016JPP}, sufficient conditions on the $c_{m,n}$ coefficients were found that when satisfied, guaranteed the convergence of the Hermite expansion in the case of additive or multiplicative separability. The resultant DFs were also demonstrated to be bounded over all velocity/momentum space. Furthermore, it was proven for certain $g_s$ function classes that the derived Hermite polynomial expansion will correspond to a non-negative DF, for at least some finite range of values of $0<\delta_s<\delta_{\text{c},s}$.


\subsection{Convergence for separable indices}\label{sec:separable}
\begin{figure}[!h]
\centering\includegraphics[width=2.5in]{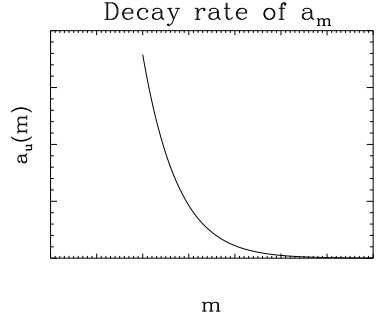}
\caption{If the modulus of the coefficients, $|a_m|$, `fall below' the graph of $(\delta_s^2 m)^{-m/2}$ as $m\to\infty$, then the Hermite series  will converge.}
\label{fig:decay}
\end{figure}

In \citet{Allanson-2016JPP}, it was proven that for $g_s$ functions compatible with $P_{zz}$ as in equation (\ref{eq:p_allanson}),
\[
g_s\propto g_{1s}(p_{x,s};\varepsilon_s)+g_{2s}(p_{y,s};\varepsilon_s),\hspace{3mm}\text{or}\hspace{3mm}g_{s}\propto g_{1s}(p_{x,s};\varepsilon_s)g_{2s}(p_{y,s};\varepsilon_s),
\] 
the corresponding Hermite expansions of the form
\begin{equation}
g_{js}(p_{js};\varepsilon_s)=\sum_{m=0}^\infty a_m\,{\rm sgn}(q_{s})^m\left(\frac{\delta_s}{\sqrt{2}}\right)^mH_{m}\left(\frac{p_{js}}{2\sqrt{\varepsilon_s}}\right)\label{eq:gj}
\end{equation}
for $j=x,y$, converge for all $p_{js}$, provided 
\begin{equation}
\lim_{m\to\infty}\sqrt{m}\left|\frac{a_{m+1}}{a_m}\right|<1/\delta_s,\label{eq:criterion}
\end{equation}
in the case of a series composed of both even- and odd-order terms, or
\begin{equation}
\lim_{m\to\infty}\,m\,\left|\frac{a_{2m+2}}{a_{2m}}\right|<1/(2\delta_s^2),\hspace{3mm}\lim_{m\to\infty}\,m\,\left|\frac{a_{2m+3}}{a_{2m+1}}\right|<1/(2\delta_s^2),\label{eq:criterion2}
\end{equation}
in the case of a series composed only of even-, or odd-order terms, respectively. In order to get a better understanding of the meaning of this theorem, it is instructive to recapitulate the results in a continuous setting. One could imagine the modulus of the coefficients, $|a_{m}|$, as a subset of the codomain of a continuous function of the independent variable $m$,
\begin{eqnarray}
&&|a_m|, \, m=0,1,2,....\nonumber\\
\to &&a=a(m),\, m\in[0,\infty), \hspace{3mm} \text{s.t.}\hspace{3mm}a(0)=|a_0|, a(1)=|a_1| ... \, .\nonumber
\end{eqnarray}
In this case, we require 
\[
a(m)=O(a_{\text{u}}(m)),\hspace{3mm}\text{s.t.}\hspace{3mm}a_u(m)=(\delta_s^2 m)^{-m/2},
\]
since the function $a_{\text{u}}$ satisfies the restrictions of Equations (\ref{eq:criterion}) and (\ref{eq:criterion2}), i.e
\begin{eqnarray}
O\left(\bigg|\frac{a_{\text{u}}(m+1)}{a_{\text{u}}(m)}\bigg|\right)&=&\frac{1}{\delta_s\sqrt{m}},\nonumber\\
O\left(\bigg|\frac{a_{\text{u}}(2m+2)}{a_{\text{u}}(2m)}\bigg|\right)&=&\frac{1}{2\delta_s^2m},\nonumber\\
O\left(\frac{a_{\text{u}}(2m+3)}{a_{\text{u}}(2m+1)}\bigg|\right)&=&\frac{1}{2\delta_s^2m}.\nonumber
\end{eqnarray}
Hence the modulus of the coefficients, $|a_m|$ must `fall below' the graph of $(\delta_s^2 m)^{-m/2}$ for large $m$, depicted in Figure \ref{fig:decay}.

\subsection{Convergence for non-separable indices}
Here we generalise the results of \citet{Allanson-2016JPP}, as detailed in Section \ref{sec:separable}, for the convergence of a Hermite expansion representation for $g_s$ that is indexed by a non-separable index, $c_{m,n}$ (i.e. the general solution that corresponds to the pressure function in equation (\ref{eq:pressureansatz})).

Essentially, the argument rests on applying the conditions for 1D indices of equations (\ref{eq:criterion}) and (\ref{eq:criterion2}), to the case of 2D indices. Formally speaking, we know that $g_s$ defined by equation (\ref{eq:gensol}), and indexed by the 2D index $c_{m,n}$, is the IPCE solution for $P_{zz}$ defined by equation (\ref{eq:pressureansatz}). Now, let $d_m$ and $e_n$ be 1D indices fixed by the following conditions, 
\begin{eqnarray}
 d_m\in D &:=&\left\{|c_{m,n^\star}| \, :\,  \lim_{m\to\infty}|c_{m+1,n^\star}/c_{m,n^\star}|=\max_{n}\left(\lim_{m\to\infty}|c_{m+1,n}/c_{m,n}|\right) , \, m=0,1,2,...\right\},\label{eq:dm}\\
  e_n\in E &:=&\left\{|c_{m^\star,n}| \, :\,  \lim_{n\to\infty}|c_{m^\star,n+1}/c_{m^\star,n}|=\max_{m}\left(\lim_{n\to\infty}|c_{m,n+1}/c_{m,n}|\right)  ,\, n=0,1,2,..\right\}.\label{eq:en}
\end{eqnarray}
That is to say that $d_m$ and $e_n$ are the 1D indices that identify the (not necessarily unique) most slowly decaying $c_{m,n}$ indices in the $m^{\text{th}}$ and $n^{\text{th}}$ `row' and `column' respectively, for large $n$ and $m$ respectively. 

Once $d_m$ and $e_n$ are identified, we can - for sufficently large $m$ and $n$ - formally bound the summand of the general solution (equation (\ref{eq:gensol})),
\[
c_{m,n}\text{sgn}(q_s)^{m+n}\left(\frac{\delta_s}{\sqrt{2}}\right)^{m+n}H_m\left(\frac{p_{x,s}}{2\sqrt{\varepsilon_s}}\right)H_n\left(\frac{p_{y,s}}{2\sqrt{\varepsilon_s}}\right)<d_{m}e_n\left(\frac{\delta_s}{\sqrt{2}}\right)^{m+n}\bigg|H_m\left(\frac{p_{x,s}}{2\sqrt{\varepsilon_s}}\right)\bigg|H_n\left(\frac{p_{y,s}}{2\sqrt{\varepsilon_s}}\right)\bigg|,
\]
and construct a sum composed of these upper bounds, according to
\begin{eqnarray}
  g_{s,\text{bound}}= \sum_{m=0}^\infty\sum_{n=0}^\infty d_{m}e_n\left(\frac{\delta_s}{\sqrt{2}}\right)^{m+n}\bigg|H_m\left(\frac{p_{x,s}}{2\sqrt{\varepsilon_s}}\right)\bigg|H_n\left(\frac{p_{y,s}}{2\sqrt{\varepsilon_s}}\right)\bigg|.\label{eq:upperbound}
\end{eqnarray}
The RHS of this equation is now of separable form. If each individual sum (over both $m$ and $n$) is convergent, then the expression on the RHS is convergent. Then, by using the comparison test (e.g. see \citet{Bartle}), convergence of the 2D series, $g_{s,\text{upper}}$, guarantees convergence of the series representation of $g_s$ in equation (\ref{eq:gensol}).

One can now treat equation (\ref{eq:upperbound}) in the same manner as in \citet{Allanson-2016JPP}, and derive conditions on the $d_m$ and $e_n$ coeeficients for convergence of the general solution, exactly analogous to those of equations (\ref{eq:criterion}) and (\ref{eq:criterion2}), and by using an upper bound on Hermite polynomials (e.g. see  \cite{Sansonebook})
\begin{equation}
| H_{j}(x)|<k\sqrt{j!}2^{j/2}\exp\left(x^2/2\right)\; \mbox{\rm s.t.\ } \; k=1.086435\, .\label{eq:hermbound}
\end{equation} 
As a result, a sufficient condition for the Hermite series representation of $g_s$ in equation (\ref{eq:gensol}) to converge is given by
\begin{equation}
\lim_{m\to\infty}\sqrt{m}\left|\frac{d_{m+1}}{d_m}\right|<1/\delta_s,\nonumber
\end{equation}
in the case of a series composed of both even- and odd-order terms, or
\begin{equation}
\lim_{m\to\infty}\,m\,\left|\frac{d_{2m+2}}{d_{2m}}\right|<1/(2\delta_s^2),\hspace{3mm}\lim_{m\to\infty}\,m\,\left|\frac{d_{2m+3}}{d_{2m+1}}\right|<1/(2\delta_s^2),\nonumber
\end{equation}
and analogously for $e_n$, with $d_m$ and $e_n$ defined by equations (\ref{eq:dm}) and (\ref{eq:en}). 

\subsection{The existence of all velocity moments}
Once the convergence of the Hermite polynomial expansion is established, then one can begin to consider the boundedness of the DF, and the existence of velocity moments. In \citet{Allanson-2016JPP}, it was shown that DFs of the form in equation (\ref{eq:gensol}) were bounded over all velocity space, but this does not guarantee that the DF has velocity space moments of all orders. For the DF to be physically meaningful, equation (\ref{eq:moments}) must be satisfied.

If the Hermite representation of $g_{s}$ is a convergent series, then by using Equation (\ref{eq:hermbound}) we see that
\[
|g_{s}(p_{x,s},p_{y,s};\varepsilon_{s})|<\mathcal{L}_{x,s}\mathcal{L}_{y,s}\exp\left(\frac{p_{x,s}^2+p_{y,s}^2}{8\varepsilon_s^2}\right) \hspace{3mm}\forall\, p_{x,s},p_{y,s}
\]
and for $\mathcal{L}_{x,s}, \mathcal{L}_{y,s}$ finite positive constants, independent of space and momentum. Now, by using the form of the DF from equation (\ref{eq:ansatz}) we see that 
\begin{eqnarray}
&&|f_s|<\exp\left[-(p_{xs}-q_sA_x)^2/(4\varepsilon_s^2)-(p_{ys}-q_sA_y)^2/(4\varepsilon_s^2)-v_z^2/(2v_{\text{th},s}^2)\right]\nonumber\\
&&\times\left(\mathcal{L}_{xs}\mathcal{L}_{ys}e^{p_{xs}^2/(8\varepsilon_s^2)}e^{p_{ys}^2/(8\varepsilon_s^2)}    \right),\nonumber
\end{eqnarray}
and we see that boundedness in momentum space (and hence velocity space) is guaranteed. The reasoning is as follows. Since $p_{js}=m_sv_j+q_sA_j$, the arguments of the exponentials scale like
\begin{equation}
\exp\left(-\frac{v_j^2}{4v_{\text{th},s}^2}\right), \label{eq:dfscale}
\end{equation}
in $v_j$ velocity space. There is also a spatial dependence in the argument of the exponential, through $A_j(z)$, but this does not affect the velocity moment at a given $z$ value. The scaling described by Expression (\ref{eq:dfscale}) not only ensures boundedness, but guarantees that velocity moments of all order exist, since
\[
\bigg|\int_{-\infty}^\infty  v^k e^{-v^2/(4v_{\text{th},s}^2)} dv \bigg|\, <\,\infty \, \forall \, k \, \in \, 0,1,2, ...
\]
Hence, a convergent Hermite series representation for the $g_s$ function, will guarantee that the DF in equation (\ref{eq:ansatz}) allows velocity moments of all orders.

\subsection{Non-negativity of the Hermite polynomial solution}
The sign of the DF as written in equation (\ref{eq:ansatz}) is identified with the sign of $g_s$, and hence non-negativity of the DF depends entirely on the non-negativity of the $g_s$ function. It was demonstrated in e.g. \citet{Channell-1976,Allanson-2016JPP}, that the non-negativity of the $P_{zz}$ function does not necessarily guarantee non-negativity of the $g_s$ function. For example, consider the pressure function originally studied by \citet{Channell-1976},
\[
P_{zz}\propto \frac{1}{2}\left(a_0+a_2\left(\frac{A_x}{B_0L}\right)^2\right)+\frac{1}{2}\left(a_0+a_2\left(\frac{A_y}{B_0L}\right)^2\right),
\]
with $a_0,a_2>0$. This pressure function is positive for all $(A_x,A_y)$. However, the corresponding $g_{s}$ function is of the form
\[
g_{s}\propto \frac{1}{2}\left[a_0+a_2\left(\frac{\delta_s}{\sqrt{2}}\right)^2H_2\left(\frac{p_{xs}}{2\sqrt{\varepsilon_s}}\right)\right]+\frac{1}{2}\left[a_0+a_2\left(\frac{\delta_s}{\sqrt{2}}\right)^2H_2\left(\frac{p_{ys}}{2\sqrt{\varepsilon_s}}\right)\right].
\]
By substituting $p_{xs}=p_{ys}=0$, we see that positivity of $g_s$ is -- for given values of $a_0$ and $a_2$ -- dependent on the size of $\varepsilon_s$, 
\begin{eqnarray}
g_s(0,0)=a_0-a_2\delta_{s}^{2},\nonumber\\
\therefore g_s(0,0)\ge 0\implies \varepsilon_s \le \frac{a_0}{a_2} \frac{e^2B_0^2L^2}{2}.\nonumber
\end{eqnarray}

It turns out that the dependence of the sign of $g_s$ on $\varepsilon_s$ (or equivalently $\delta_s$), seems to be a rather general principle. In \citet{Allanson-2016JPP} it was proven that for a smooth $P_{zz}$ function (and either summatively or multiplicatively separable), and under the \emph{a priori} assumption of a continuous $g_s$ function that is uniformly bounded from below in momentum space, the corresponding $g_s$ function is non-negative for at least a finite range of $\delta_s$ values, i.e. for all $\delta_{s}\le \delta_{\text{c},s}$, for $\delta_{\text{c},s}\in (0,\infty)$ some critical value of $\delta_{s}$, and as yet undetermined. Here we generalise this proof for the arbitrarily indexed general solution of the form in equation (\ref{eq:gensol}).

\subsection{The limit as $\delta_s \to 0$, $B_0L\to\infty$ and fixed $\varepsilon_s$}
First suppose that for a given value of $\delta_s$, that there exists some regions in $(p_{x,s},p_{y,s})$ space where $g_s<0$. Then, the statement that $g_{s}$ has a finite lower bound, combined with the expression in Equation (\ref{eq:gensol}) implies that the $g_s$ function is bounded below by a finite constant of the form $c_{0,0}+\delta_s \mathcal{M}$, with
\[
\mathcal{M}=\frac{1}{\sqrt{2}}\inf_{(p_{x,s},p_{y,s})}\sum_{n=1}^\infty \sum_{m=1}^\infty c_{m,n}\text{sgn}(q_s)^{m+n}\left(\frac{\delta_s}{\sqrt{2}}\right)^{m+n-1}H_m\left(\frac{p_{x,s}}{2\sqrt{\varepsilon_s}}\right)H_n\left(\frac{p_{y,s}}{2\sqrt{\varepsilon_s}}\right),
\]
and finite, i.e. the greatest lower bound). In \citet{Allanson-2016JPP}, the next step in the (similar) proof was to let $\delta_s\to 0$, independently of $\varepsilon_s$. We can see from equation (\ref{eq:tuning}), that this is equivalent to sending $B_0L\to\infty$. We see that by letting $\delta_s\to 0$ 
\[
\lim_{\delta_s\to 0} g_{s}=c_{0,0}=\lim_{B_0L\to\infty}P_{zz}\ge 0,
\]
since $P_{zz}\ge 0\, \forall (A_x,A_y)$. Therefore, there must exist some critical value of $\delta_s=\delta_{s,c}$, such that for all $\delta_s<\delta_c$, $g_s$ is non-negative. Note that if the negative patches of $g_{js}$ do not exist for any $\delta_s$, then trivially $\delta_{s,c}=\infty$ as a special case.

\subsection{The limit as $\delta_s\to 0$, $\varepsilon_s \to 0$, and fixed $B_0L$}
This result can also be shown by considering $B_0L$ as fixed parameters, and sending $v_{\text{th},s}\to 0$ (i.e. $\delta_s\to 0$ and $\varepsilon_s \to 0$), as follows. Using the fact that a $j^{\text{th}}$ order Hermite polynomial, $H_j(X)$, is a $j^{\text{th}}$ order polynomial, with highest order term of the form
\[
2^jX^j,
\]
it can be seen from equations (\ref{eq:gensol}) and (\ref{eq:tuning}) that, for fixed $B_0L$, 
\begin{eqnarray}
\lim_{\varepsilon_s\to 0} g_{s}(p_{x,s},p_{y,s})&=&\sum_{m,n}c_{m,n}\left(\frac{\text{sgn}(q_s)m_sv_{\text{th},s}}{\sqrt{2}eB_0L}\right)^{m+n}\left(\frac{p_{x,s}}{\sqrt{\varepsilon_s}}\right)^m\left(\frac{p_{y,s}}{\sqrt{\varepsilon_s}}\right)^n\nonumber ,\\  
&=&\sum_{m,n}c_{m,n}\left(\frac{p_{x,s}}{q_sB_0L}\right)^m\left(\frac{p_{y,s}}{q_sB_0L}\right)^n\nonumber ,\\ 
&=&n_{0}^{-1}\frac{\beta_e\beta_i}{\beta_e+\beta_i}P_{zz}\left(\frac{p_{x,s}}{q_s},\frac{p_{y,s}}{q_s}\right).
\end{eqnarray}
We can write down the first equality since only the highest order terms survive in each of the individual Hermite polynomials. The second equality follows from algebraic manipulation, and the final equality follows from equation (\ref{eq:pressureansatz}). The RHS of the third equality is non-negative, since $P_{zz}\ge 0\, \forall (A_x,A_y)$. Hence, we see that $g_s$ function can be written as 
\begin{equation}
g_s(p_{x,s},p_{y,s})=P_{zz}\left(\frac{p_{x,s}}{q_s},\frac{p_{y,s}}{q_s}\right)+\varepsilon_s\mathcal{N},\label{eq:newinf}
\end{equation}
for $\varepsilon_s\mathcal{N}$ equal to the sum of the terms in the Hermite expansion of equation (\ref{eq:gensol}), except those that correspond to the highest order term of each Hermite polynomial. Using equation (\ref{eq:newinf}), and the assumption that $g_s$ is uniformly bounded from below, we see that $g_s$ converges uniformly to a positve function, $P_{zz}\left(\frac{p_{x,s}}{q_s},\frac{p_{y,s}}{q_s}\right)$, as $\varepsilon_s\to 0$, and for a fixed $B_0L$. Hence, using similar logic to above, there must exist some critical value of $\varepsilon_s=\varepsilon_{s,c}$, such that for all $\varepsilon_s<\varepsilon_{s,c}$, $g_s$ is non-negative. Note that if the negative patches of $g_{s}$ do not exist for any $\varepsilon_s$, then trivially $\delta_{s,c}=\infty$ as a special case.

\section{Discussion \& Conclusions}
This paper has introduced and reviewed the theory first of collisionless plasma equilibria (Vlasov-Maxwell equilibria), and then of the inverse problem in collisionless plasma equilibria (IPCE) in a general sense. Then we have applied this theory to the case of one-dimensional and strictly neutral magnetised plasmas. We have demonstrated that in this context, IPCE can reduce to the inversion of Weierstrass transforms, and discussed the parallels between IPCE and `backwards solutions of the heat equation'. It will be a very interesting topic for future investigation to see if this analogy can bring further useful insight and results. 

The main theoretical developments of this paper have focussed on the mathematical criteria that a candidate solution of IPCE must satisfy, and in particular for those solutions obtained by use of a Hermite polynomial expansion. We have reviewed the recent work by \citet{Allanson-2016JPP} on this topic, and presented a rigorous treatment of the use of Hermite polynomial expansions for IPCE as applied to 1D strictly neutral magnetised plasmas. We have derived new results relating to convergence and non-negativity of a candidate solution for IPCE, as well as the existence of velocity moments of all orders, for distribution functions that are consistent with an arbitrarily indexed 2D Maclaurin expansion of the pressure function. In particular, we have proven that non-negative solutions of IPCE will exist over all momentum space, and for some sections of parameter space, for candidate solutions belonging to a certain class. Future work should focus on extending the results regarding non-negativity to a broader class of solutions, since at present we have \emph{a priori} assumed that the naive/formal solution to IPCE is uniformly bounded from below, over all parameter and momentum space. It would be useful to understand to what extent this condition can be relaxed, and whether the aforementioned analogy with the heat equation can be brought to bear on this problem. Furthermore, we would like to establish precisely over which values of parameter space the candidate solution is non-negative, i.e. extend the results from being purely \emph{existence} results, to something more concrete.

The other obvious generalisation is to relax earlier assumptions relating to the macroscopic nature of the plasma. For example, to what extent does IPCE change when applied to spatially 2D plasmas, non-neutral plasmas, non-planar (e.g. cylindrical) geometries, or even the inclusion of gravitational effects? Future work could be directed in these directions, motivated by the many possible applications in plasma physics.

\section*{Acknowledgment}
OA gratefully acknowledges the financial support of the Science and Technology Facilities Council Consolidated Grant Nos. ST/K000950/1 and ST/N000609/1, the Science and Technology Facilities Council Doctoral Training Grant No. ST/K502327/1, and the Natural Environment Research Council Grant No. NE/P017274/1 (Rad-Sat). ST gratefully acknowledges the financial support of Engineering and Physical Sciences Research Council Doctoral Training Grant No. EP/K503162/1, Natural Sciences and Engineering Research Council of Canada Grant Nos. 2016-03719 and RGPIN-2014-03154, and the Faculty of Mathematics, University of Waterloo.


%


\bibliographystyle{imamat}

\begin{thebibliography}{}

\bibitem[{Abraham-Shrauner}, 1968]{Abraham-Shrauner-1968}
{Abraham-Shrauner}, B. (1968)  {Exact, Stationary Wave Solutions of the
  Nonlinear Vlasov Equation}. {\em Physics of Fluids}, \textbf{11}, 1162--1167.

\bibitem[{Abraham-Shrauner}, 2013]{Abraham-Shrauner-2013}
{Abraham-Shrauner}, B. (2013)  {Force-free Jacobian equilibria for
  Vlasov-Maxwell plasmas}. {\em Physics of Plasmas}, \textbf{20}(10), 102117.

\bibitem[{Allanson} et~al., 2016]{Allanson-2016JPP}
{Allanson}, O., {Neukirch}, T., {Troscheit}, S. {\&} {Wilson}, F. (2016)  {From
  one-dimensional fields to Vlasov equilibria: theory and application of
  Hermite polynomials}. {\em Journal of Plasma Physics}, \textbf{82}(3),
  905820306.

\bibitem[{Allanson} et~al., 2015]{Allanson-2015POP}
{Allanson}, O., {Neukirch}, T., {Wilson}, F. {\&} {Troscheit}, S. (2015)  {An
  exact collisionless equilibrium for the Force-Free Harris Sheet with low
  plasma beta}. {\em Physics of Plasmas}, \textbf{22}(10), 102116.

\bibitem[{Alpers}, 1969]{Alpers-1969}
{Alpers}, W. (1969)  {Steady State Charge Neutral Models of the Magnetopause}.
  {\em Astrophysics and Space Science}, \textbf{5}, 425--437.

\bibitem[{Arfken} \& {Weber}, 2001]{Arfkenbook}
{Arfken}, G.~B. {\&} {Weber}, H.~J. (2001) {\em Mathematical methods for
  physicists}.
Harcourt/Academic Press, Burlington, MA, fifth edition.

\bibitem[{Attico} \& {Pegoraro}, 1999]{Attico-1999}
{Attico}, N. {\&} {Pegoraro}, F. (1999)  {Periodic equilibria of the
  Vlasov-Maxwell system}. {\em Physics of Plasmas}, \textbf{6}, 767--770.

\bibitem[{Aunai} et~al., 2013]{Aunai-2013}
{Aunai}, N., {Hesse}, M., {Zenitani}, S., {Kuznetsova}, M., {Black}, C.,
  {Evans}, R. {\&} {Smets}, R. (2013)  {Comparison between hybrid and fully
  kinetic models of asymmetric magnetic reconnection: Coplanar and guide field
  configurations}. {\em Physics of Plasmas}, \textbf{20}(2), 022902.

\bibitem[Bartle \& Sherbert, 2000]{Bartle}
Bartle, R. {\&} Sherbert, D. (2000) {\em Introduction to real analysis}.
John Wiley \& Sons Canada, Limited.

\bibitem[Baumjohann \& Treumann, 1997]{Baumjohannbook}
Baumjohann, W. {\&} Treumann, R. (1997) {\em Basic Space Plasma Physics}.
Imperial College Press.

\bibitem[{Belmont} et~al., 2012]{Belmont-2012}
{Belmont}, G., {Aunai}, N. {\&} {Smets}, R. (2012)  {Kinetic equilibrium for an
  asymmetric tangential layer}. {\em Physics of Plasmas}, \textbf{19}(2),
  022108.

\bibitem[Belmont et~al., 2013]{belmontbook}
Belmont, G., Grappin, R., Mottez, F., Pantellini, F. {\&} Pelletier, G. (2013)
  {\em Collisionless Plasmas in Astrophysics}.
Wiley.

\bibitem[Bennett, 1934]{Bennett-1934}
Bennett, W.~H. (1934)  Magnetically Self-Focussing Streams. {\em Physical
  Review}, \textbf{45}(12), 890--897.

\bibitem[{Bernstein} et~al., 1957]{Bernstein-1957}
{Bernstein}, I.~B., {Greene}, J.~M. {\&} {Kruskal}, M.~D. (1957)  {Exact
  Nonlinear Plasma Oscillations}. {\em Physical Review}, \textbf{108},
  546--550.

\bibitem[Bilodeau, 1962]{Bilodeau-1962}
Bilodeau, G.~G. (1962)  The {Weierstrass} transform and {Hermite} polynomials.
  {\em Duke Mathematical Journal}, \textbf{29}(2), 293--308.

\bibitem[Birdsall \& Langdon, 2004]{birdsallbook}
Birdsall, C. {\&} Langdon, A. (2004) {\em Plasma Physics via Computer
  Simulation}.
Series in Plasma Physics and Fluid Dynamics. Taylor \& Francis.

\bibitem[{Bobrova} et~al., 2001]{Bobrova-2001}
{Bobrova}, N.~A., {Bulanov}, S.~V., {Sakai}, J.~I. {\&} {Sugiyama}, D. (2001)
  {Force-free equilibria and reconnection of the magnetic field lines in
  collisionless plasma configurations}. {\em Physics of Plasmas}, \textbf{8},
  759--768.

\bibitem[{Bobrova} \& {Syrovatski{\v i}}, 1979]{Bobrova-1979}
{Bobrova}, N.~A. {\&} {Syrovatski{\v i}}, S.~I. (1979)  {Violent instability of
  one-dimensional forceless magnetic field in a rarefied plasma}. {\em Soviet
  Journal of Experimental and Theoretical Physics Letters}, \textbf{30},
  535--+.

\bibitem[Burgess \& Scholer, 2015]{Burgessbook}
Burgess, D. {\&} Scholer, M. (2015) {\em Collisionless Shocks in Space Plasmas:
  Structure and Accelerated Particles}.
Cambridge Atmospheric and Space Science Series. Cambridge University Press.

\bibitem[{Camporeale} et~al., 2006]{Camporeale-2006}
{Camporeale}, E., {Delzanno}, G.~L., {Lapenta}, G. {\&} {Daughton}, W. (2006)
  {New approach for the study of linear Vlasov stability of inhomogeneous
  systems}. {\em Physics of Plasmas}, \textbf{13}(9), 092110.

\bibitem[{Channell}, 1976]{Channell-1976}
{Channell}, P.~J. (1976)  {Exact Vlasov-Maxwell equilibria with sheared
  magnetic fields}. {\em Physics of Fluids}, \textbf{19}, 1541--1545.

\bibitem[Chen, 2015]{chenbook}
Chen, F. (2015) {\em Introduction to Plasma Physics and Controlled Fusion}.
Springer International Publishing.

\bibitem[{Coppi} et~al., 1966]{Coppi-1966}
{Coppi}, B., {Laval}, G. {\&} {Pellat}, R. (1966)  {Dynamics of the Geomagnetic
  Tail}. {\em Physical Review Letters}, \textbf{16}, 1207--1210.

\bibitem[{Daughton}, 1999]{Daughton-1999}
{Daughton}, W. (1999)  {The unstable eigenmodes of a neutral sheet}. {\em
  Physics of Plasmas}, \textbf{6}, 1329--1343.

\bibitem[{Drake} \& {Lee}, 1977]{Drake-1977}
{Drake}, J.~F. {\&} {Lee}, Y.~C. (1977)  {Kinetic theory of tearing
  instabilities}. {\em Physics of Fluids}, \textbf{20}, 1341--1353.

\bibitem[Evans, 2010]{Evansbook}
Evans, L.~C. (2010) {\em Partial differential equations}, volume~19 of {\em
  Graduate Studies in Mathematics}.
American Mathematical Society, Providence, RI, second edition.

\bibitem[Fitzpatrick, 2014]{Fitzpatrickbook}
Fitzpatrick, R. (2014) {\em Plasma Physics: An Introduction}.
CRC Press, Taylor \& Francis Group.

\bibitem[{Freidberg}, 1987]{Freidbergbook}
{Freidberg}, J.~P. (1987) {\em Ideal Magnetohydrodynamics}.
Plenum Publishing Corportation.

\bibitem[{Fu} \& {Hau}, 2005]{Fu-2005}
{Fu}, W.-Z. {\&} {Hau}, L.-N. (2005)  {Vlasov-Maxwell equilibrium solutions for
  Harris sheet magnetic field with Kappa velocity distribution}. {\em Physics
  of Plasmas}, \textbf{12}(7), 070701--+.

\bibitem[Gary, 2005]{Gary-2005}
Gary, S. (2005) {\em Theory of Space Plasma Microinstabilities}.
Cambridge Atmospheric and Space Science Series. Cambridge University Press.

\bibitem[{Ghosh} et~al., 2014]{Ghosh-2014}
{Ghosh}, A., {Janaki}, M.~S., {Dasgupta}, B. {\&} {Bandyopadhyay}, A. (2014)
  {Chaotic magnetic fields in Vlasov-Maxwell equilibria}. {\em Chaos},
  \textbf{24}(1), 013117.

\bibitem[Grad, 1949a]{Grad-1949a}
Grad, H. (1949a)  Note on {$N$}-dimensional {H}ermite polynomials. {\em Comm.
  Pure Appl. Math.}, \textbf{2}, 325--330.

\bibitem[Grad, 1949b]{Grad-1949b}
Grad, H. (1949b)  On the kinetic theory of rarefied gases. {\em Comm. Pure
  Appl. Math.}, \textbf{2}, 331--407.

\bibitem[{Grad}, 1961]{Grad-1961}
{Grad}, H. (1961)  {Boundary Layer between a Plasma and a Magnetic Field}. {\em
  Physics of Fluids}, \textbf{4}, 1366--1375.

\bibitem[{Greene}, 1993]{Greene-1993}
{Greene}, J.~M. (1993)  {One-dimensional Vlasov-Maxwell equilibria}. {\em
  Physics of Fluids B}, \textbf{5}, 1715--1722.

\bibitem[Guo et~al., 2014]{Guo-2014}
Guo, F., Li, H., Daughton, W. {\&} Liu, Y.-H. (2014)  Formation of Hard Power
  Laws in the Energetic Particle Spectra Resulting from Relativistic Magnetic
  Reconnection. {\em Phys. Rev. Lett.}, \textbf{113}, 155005.

\bibitem[{Harris}, 1962]{Harris-1962}
{Harris}, E.~G. (1962)  {On a plasma sheath separating regions of oppositely
  directed magnetic field}. {\em Nuovo Cimento}, \textbf{23}, 115.

\bibitem[{Harrison} \& {Neukirch}, 2009a]{Harrison-2009PRL}
{Harrison}, M.~G. {\&} {Neukirch}, T. (2009a)  {One-Dimensional Vlasov-Maxwell
  Equilibrium for the Force-Free Harris Sheet}. {\em Physical Review Letters},
  \textbf{102}(13), 135003--+.

\bibitem[{Harrison} \& {Neukirch}, 2009b]{Harrison-2009POP}
{Harrison}, M.~G. {\&} {Neukirch}, T. (2009b)  {Some remarks on one-dimensional
  force-free Vlasov-Maxwell equilibria}. {\em Physics of Plasmas},
  \textbf{16}(2), \\ 022106--+.

\bibitem[{Henon}, 1982]{Henon-1982}
{Henon}, M. (1982)  {Vlasov equation}. {\em Astronomy and Astrophysics},
  \textbf{114}, 211.

\bibitem[Hesse et~al., 2016]{Hesse-2016}
Hesse, M., Aunai, N., Birn, J., Cassak, P., Denton, R.~E., Drake, J.~F.,
  Gombosi, T., Hoshino, M., Matthaeus, W., Sibeck, D. {\&} Zenitani, S. (2016)
  Theory and Modeling for the Magnetospheric Multiscale Mission. {\em Space
  Science Reviews}, \textbf{199}(1), 577--630.

\bibitem[{Hesse} et~al., 2014]{Hesse-2014}
{Hesse}, M., {Aunai}, N., {Sibeck}, D. {\&} {Birn}, J. (2014)  {On the electron
  diffusion region in planar, asymmetric, systems}. {\em Geophysical Research
  Letters}, \textbf{41}, 8673--8680.

\bibitem[Hesse et~al., 2013]{Hesse-2013}
Hesse, M., Aunai, N., Zenitani, S., Kuznetsova, M. {\&} Birn, J. (2013)
  Aspects of collisionless magnetic reconnection in asymmetric systems. {\em
  Physics of Plasmas}, \textbf{20}(6).

\bibitem[{Hesse} et~al., 2005]{Hesse-2005}
{Hesse}, M., {Kuznetsova}, M., {Schindler}, K. {\&} {Birn}, J. (2005)
  {Three-dimensional modeling of electron quasiviscous dissipation in
  guide-field magnetic reconnection}. {\em Physics of Plasmas},
  \textbf{12}(10), 100704--+.

\bibitem[{Hewett} et~al., 1976]{Hewett-1976}
{Hewett}, D.~W., {Nielson}, C.~W. {\&} {Winske}, D. (1976)  {Vlasov confinement
  equilibria in one dimension}. {\em Physics of Fluids}, \textbf{19}, 443--449.

\bibitem[{Hurley}, 1963]{Hurley-1963}
{Hurley}, J. (1963)  {Analysis of the Transition Region between a Uniform
  Plasma and its Confining Magnetic Field. II}. {\em Physics of Fluids},
  \textbf{6}, 83--88.

\bibitem[{Janaki} \& {Dasgupta}, 2012]{Janaki-2012}
{Janaki}, M.~S. {\&} {Dasgupta}, B. (2012)  {Vlasov-Maxwell equilibria:
  Examples from higher-curl Beltrami magnetic fields}. {\em Physics of
  Plasmas}, \textbf{19}(3), 032113.

\bibitem[{Jeans}, 1915]{Jeans-1915}
{Jeans}, J.~H. (1915)  {On the theory of star-streaming and the structure of
  the universe}. {\em Monthly Notices of the Royal Astronomical Society},
  \textbf{76}, 70--84.

\bibitem[John, 1991]{johnbook}
John, F. (1991) {\em Partial Differential Equations}.
Applied Mathematical Sciences. Springer New York.

\bibitem[{Kocharovsky} et~al., 2010]{Kocharovsky-2010}
{Kocharovsky}, V.~V., {Kocharovsky}, V.~V. {\&} {Martyanov}, V.~J. (2010)
  {Self-Consistent Current Sheets and Filaments in Relativistic Collisionless
  Plasma with Arbitrary Energy Distribution of Particles}. {\em Physical Review
  Letters}, \textbf{104}(21), 215002.

\bibitem[{Kolotkov} et~al., 2015]{Kolotkov-2015}
{Kolotkov}, D.~Y., {Vasko}, I.~Y. {\&} {Nakariakov}, V.~M. (2015)  {Kinetic
  model of force-free current sheets with non-uniform temperature}. {\em
  Physics of Plasmas}, \textbf{22}(11), 112902.

\bibitem[{Krall} \& {Trivelpiece}, 1973]{krallbook-1973}
{Krall}, N.~A. {\&} {Trivelpiece}, A.~W. (1973) {\em {Principles of plasma
  physics}}.
International Student Edition - International Series in Pure and Applied
  Physics, Tokyo: McGraw-Hill Kogakusha.

\bibitem[Kulsrud, 2005]{kulsrudbook}
Kulsrud, R. (2005) {\em Plasma Physics for Astrophysics}.
Plasma Physics for Astrophysics. Princeton University Press.

\bibitem[{Kulsrud}, 1983]{KulsrudMHD}
{Kulsrud}, R.~M. (1983) {\em {MHD description of plasma. In Handbook of Plasma
  Physics, Volume 1}}.
Amsterdam: North-Holland.

\bibitem[{Lakhina} \& {Schindler}, 1983]{Lakhina-1983}
{Lakhina}, G.~S. {\&} {Schindler}, K. (1983)  {Tearing modes in the
  magnetopause current sheet}. {\em Astrophysics and Space Science},
  \textbf{97}, 421--426.

\bibitem[Landau, 1946]{Landau-1946}
Landau, L. (1946)  {On the vibrations of the electronic plasma}. {\em
  J.Phys.(USSR)}, \textbf{10}, 25--34.

\bibitem[{Landau} \& {Lifshitz}, 2013]{Landaufields}
{Landau}, L.~D. {\&} {Lifshitz}, E.~M. (2013) {\em The Classical Theory of
  Fields}.
Course of Theoretical Physics. Elsevier Science.

\bibitem[{Lee} \& {Kan}, 1979]{Lee-1979JGR}
{Lee}, L.~C. {\&} {Kan}, J.~R. (1979)  {A unified kinetic model of the
  tangential magnetopause structure}. {\em Journal of Geophysical Research
  (Space Physics)}, \textbf{84}, 6417--6426.

\bibitem[{Lerche}, 1967]{Lerche-1967}
{Lerche}, I. (1967)  {On the Boundary Layer between a Warm, Streaming Plasma
  and a Confined Magnetic Field}. {\em Journal of Geophysical Research (Space
  Physics)}, \textbf{72}, 5295--+.

\bibitem[Lifshitz \& Pitaevski{\u\i}, 1981]{lifshitzkinetic}
Lifshitz, E. {\&} Pitaevski{\u\i}, L. (1981) {\em Physical kinetics}.
Course of theoretical physics. Butterworth-Heinemann.

\bibitem[{Liu} \& {Hesse}, 2016]{Liu-2016}
{Liu}, Y.-H. {\&} {Hesse}, M. (2016)  {Suppression of collisionless magnetic
  reconnection in asymmetric current sheets}. {\em Physics of Plasmas},
  \textbf{23}(6), 060704.

\bibitem[{Lynden-Bell}, 1962]{Lynden-Bell-1962}
{Lynden-Bell}, D. (1962)  {Stellar dynamics: Exact solution of the
  self-gravitation equation}. {\em Monthly Notices of the Royal Astronomical
  Society}, \textbf{123}, 447.

\bibitem[{Malakit} et~al., 2010]{Malakit-2010}
{Malakit}, K., {Shay}, M.~A., {Cassak}, P.~A. {\&} {Bard}, C. (2010)  {\\
  Scaling of asymmetric magnetic reconnection: Kinetic particle-in-cell
  simulations}. {\em Journal of Geophysical Research (Space Physics)},
  \textbf{115}, A10223.

\bibitem[Marsch, 2006]{Marsh-2006}
Marsch, E. (2006)  Kinetic Physics of the Solar Corona and Solar Wind. {\em
  Living Reviews in Solar Physics}, \textbf{3}(1).

\bibitem[{Marsh}, 1996]{Marshbook}
{Marsh}, G. (1996) {\em Force-Free Magnetic Fields: Solutions, Topology and
  Applications}.
World Scientific, Singapore.

\bibitem[{Montgomery} \& {Joyce}, 1969]{Montgomery-1969}
{Montgomery}, D. {\&} {Joyce}, G. (1969)  {Shock-like solutions of the
  electrostatic Vlasov equation}. {\em Journal of Plasma Physics}, \textbf{3},
  1--11.

\bibitem[{Moratz} \& {Richter}, 1966]{Moratz-1966}
{Moratz}, E. {\&} {Richter}, E.~W. (1966)
  {\\Elektronen-Geschwindigkeitsverteilungsfunktionen f{\"u}r kraftfreie bzw.
  teilweise kraftfreie Magnetfelder}. {\em Zeitschrift Naturforschung Teil A},
  \textbf{21}, 1963.

\bibitem[Mouhot \& Villani, 2011]{Mouhot-2011}
Mouhot, C. {\&} Villani, C. (2011)  On {L}andau damping. {\em Acta Math.},
  \textbf{207}(1), 29--201.

\bibitem[{Mynick} et~al., 1979]{Mynick-1979a}
{Mynick}, H.~E., {Sharp}, W.~M. {\&} {Kaufman}, A.~N. (1979)  {Realistic Vlasov
  slab equilibria with magnetic shear}. {\em Physics of Fluids}, \textbf{22},
  1478--1484.

\bibitem[{Neukirch} et~al., 2009]{Neukirch-2009}
{Neukirch}, T., {Wilson}, F. {\&} {Harrison}, M.~G. (2009)  {A detailed
  investigation of the properties of a Vlasov-Maxwell equilibrium for the
  force-free Harris sheet}. {\em Physics of Plasmas}, \textbf{16}(12), 122102.

\bibitem[{Nicholson}, 1963]{Nicholson-1963}
{Nicholson}, R.~B. (1963)  {Solution of the Vlasov Equations for a Plasma in an
  Externally Uniform Magnetic Field}. {\em Physics of Fluids}, \textbf{6},
  1581--1586.

\bibitem[{Panov} et~al., 2011]{Panov-2011}
{Panov}, E.~V., {Artemyev}, A.~V., {Nakamura}, R. {\&} {Baumjohann}, W. (2011)
  {Two types of tangential magnetopause current sheets: Cluster observations
  and theory}. {\em Journal of Geophysical Research (Space Physics)},
  \textbf{116}, A12204.

\bibitem[{Pegoraro} et~al., 2015]{Pegoraro-2015}
{Pegoraro}, F., {Califano}, F., {Manfredi}, G. {\&} {Morrison}, P.~J. (2015)
  {Theory and applications of the Vlasov equation}. {\em European Physical
  Journal D}, \textbf{69}, 68.

\bibitem[{Pritchett}, 2008]{Pritchett-2008}
{Pritchett}, P.~L. (2008)  {Collisionless magnetic reconnection in an
  asymmetric current sheet}. {\em Journal of Geophysical Research (Space
  Physics)}, \textbf{113}, A06210.

\bibitem[{Quest} \& {Coroniti}, 1981]{Quest-1981A}
{Quest}, K.~B. {\&} {Coroniti}, F.~V. (1981)  {Linear theory of tearing in a
  high-beta plasma}. {\em Journal of Geophysical Research}, \textbf{86},
  3299--3305.

\bibitem[Sansone, 1959]{Sansonebook}
Sansone, G. (1959) {\em Orthogonal functions}.
Revised English ed. Translated from the Italian by A. H. Diamond; with a
  foreword by E. Hille. Pure and Applied Mathematics, Vol. IX. Interscience
  Publishers, Inc., New York; Interscience Publishers, Ltd., London.

\bibitem[Sauvigny, 2012]{sauvignybook}
Sauvigny, F. (2012) {\em Partial Differential Equations 1: Foundations and
  Integral Representations}.
Universitext. Springer London.

\bibitem[{Schamel}, 1971]{Schamel-1971}
{Schamel}, H. (1971)  {Stationary solutions of the electrostatic Vlasov
  equation}. {\em Plasma Physics}, \textbf{13}, 491--505.

\bibitem[{Schamel}, 1972]{Schamel-1972JPP}
{Schamel}, H. (1972)  {Non-linear electrostatic plasma waves}. {\em Journal of
  Plasma Physics}, \textbf{7}, 1--12.

\bibitem[{Schamel}, 1986]{Schamel-1986}
{Schamel}, H. (1986)  {Electron holes, ion holes and double layers.
  Electrostatic phase space structures in theory and experiment}. {\em Physics
  Reports}, \textbf{140}, 161--191.

\bibitem[Schekochihin et~al., 2016]{Schekochihin-2016}
Schekochihin, A.~A., Parker, J.~T., Highcock, E.~G., Dellar, P.~J., Dorland, W.
  {\&} Hammett, G.~W. (2016)  Phase mixing versus nonlinear advection in
  drift-kinetic plasma turbulence. {\em Journal of Plasma Physics},
  \textbf{82}, \\ 905820212 (47 pages).

\bibitem[Schindler, 2007]{Schindlerbook}
Schindler, K. (2007) {\em Physics of Space Plasma Activity}.
Cambridge University Press.

\bibitem[Schmid-Burgk, 1965]{Schmid-Burgk-1965}
Schmid-Burgk, J. (1965)  Zweidimensionale selbstkonsistente L\"{o}sungen der
  station\"{a}ren Wlassowgleichung f\"{u}r Zweikomponentplasmen. \\
  Max-Planck-Institut f\"{u}r Physik und Astrophysik, Master's thesis.

\bibitem[{Sestero}, 1964]{Sestero-1964}
{Sestero}, A. (1964)  {Structure of Plasma Sheaths}. {\em Physics of Fluids},
  \textbf{7}, 44--51.

\bibitem[{Sestero}, 1965]{Sestero-1965}
{Sestero}, A. (1965)  {Charge Separation Effects in the Ferraro-Rosenbluth Cold
  Plasma Sheath Model}. {\em Physics of Fluids}, \textbf{8}, 739--744.

\bibitem[{Suzuki} \& {Shigeyama}, 2008]{Suzuki-2008}
{Suzuki}, A. {\&} {Shigeyama}, T. (2008)  {A novel method to construct
  stationary solutions of the Vlasov-Maxwell system}. {\em Physics of Plasmas},
  \textbf{15}(4), 042107--+.

\bibitem[{Swisdak} et~al., 2003]{Swisdak-2003}
{Swisdak}, M., {Rogers}, B.~N., {Drake}, J.~F. {\&} {Shay}, M.~A. (2003)
  {Diamagnetic suppression of component magnetic reconnection at the
  magnetopause}. {\em Journal of Geophysical Research (Space Physics)},
  \textbf{108}, 1218.

\bibitem[{Tassi} et~al., 2008]{Tassi-2008}
{Tassi}, E., {Pegoraro}, F. {\&} {Cicogna}, G. (2008)  {Solutions and
  symmetries of force-free magnetic fields}. {\em Physics of Plasmas},
  \textbf{15}(9), 092113--+.

\bibitem[Tasso, 1969]{Tasso-1969}
Tasso, H. (1969)  Non-linear quasi-neutral electrostatic plasma waves and shock
  waves. {\em Plasma Physics}, \textbf{11}(8), 663.

\bibitem[Vaivads et~al., 2016]{Vaivads-2016}
Vaivads, A., Retinò, A., Soucek, J., Khotyaintsev, Y.~V., Valentini, F.,
  Escoubet, C.~P., Alexandrova, O., André, M., Bale, S.~D., Balikhin, M. {\&}
  et~al. (2016)  Turbulence Heating ObserveR - satellite mission proposal. {\em
  Journal of Plasma Physics}, \textbf{82}(5).

\bibitem[{Vasko} et~al., 2013]{Vasko-2013}
{Vasko}, I.~Y., {Artemyev}, A.~V., {Popov}, V.~Y. {\&} {Malova}, H.~V. (2013)
  {Kinetic models of two-dimensional plane and axially symmetric current
  sheets: Group theory approach}. {\em Physics of Plasmas}, \textbf{20}(2),
  022110.

\bibitem[Vlasov, 1968]{Vlasov-1968}
Vlasov, A.~A. (1968)  The vibrational properties of an electron gas. {\em
  Physics-Uspekhi}, \textbf{10}(6), 721--733.

\bibitem[Weinberg, 2005]{Weinbergbook}
Weinberg, S. (2005) {\em The quantum theory of fields. {V}ol. {I}}.
Cambridge University Press, Cambridge.
Foundations.

\bibitem[Widder, 1951]{Widder-1951}
Widder, D.~V. (1951)  {Necessary and sufficient conditions for the
  representation of a function by a Weierstrass transform}. {\em Transactions
  of the American Mathematical Society}, \textbf{71}, 430--439.

\bibitem[Widder, 1954]{Widder-1954}
Widder, D.~V. (1954)  The convolution transform. {\em Bulletin of the American
  Mathematical Society}, \textbf{60}(5), 444--456.

\bibitem[{Wilson} \& {Neukirch}, 2011]{Wilson-2011}
{Wilson}, F. {\&} {Neukirch}, T. (2011)  {A family of one-dimensional
  Vlasov-Maxwell equilibria for the force-free Harris sheet}. {\em Physics of
  Plasmas}, \textbf{18}(8), 082108.

\bibitem[{Yoon} \& {Lui}, 2005]{Yoon-2005}
{Yoon}, P.~H. {\&} {Lui}, A.~T.~Y. (2005)  {A class of exact two-dimensional
  kinetic current sheet equilibria}. {\em Journal of Geophysical Research
  (Space Physics)}, \textbf{110}, 1202--+.

\bibitem[{Zayed}, 1996]{zayedbook}
{Zayed}, A.~I. (1996) {\em Handbook of function and generalized function
  transformations}.
Mathematical Sciences Reference Series. CRC Press, Boca Raton, FL.

\bibitem[{Zocco}, 2015]{Zocco-2015}
{Zocco}, A. (2015)  {Linear collisionless Landau damping in Hilbert space}.
  {\em Journal of Plasma Physics}, \textbf{81}(4), 049002.

\end{thebibliography}


\end{document}